\documentclass[aps,twocolumn,pra,showpacs,amsmath,amssymb,showkeys,10pt]{revtex4-1}

\usepackage{graphicx}
\usepackage{epstopdf}
\usepackage{braket}
\usepackage{hyperref}

\begin{document}

\title{Monitoring the resonantly driven Jaynes-Cummings oscillator by an external two-level emitter: A cascaded open-systems approach}

\author{Th. K. Mavrogordatos}
\email[Email address: ]{themis.mavrogordatos@fysik.su.se}
\affiliation{Department of Physics, Stockholm University, SE-106 91, Stockholm, Sweden}

\author{J. Larson}
\affiliation{Department of Physics, Stockholm University, SE-106 91, Stockholm, Sweden}

\date{\today}

\begin{abstract}
We address the consequences of backaction in the unidirectional coupling of two cascaded open quantum subsystems connected to the same reservoir at different spatial locations. In the spirit of [H. J. Carmichael, Phys. Rev. Lett. {\bf 70}, 2273 (1993)], the second subsystem is a two-level atom, while the first transforms from a driven empty cavity to a perturbative QED configuration and ultimately to a driven Jaynes-Cummings (JC) oscillator through a varying light-matter coupling strength. For our purpose, we appeal at first to the properties of resonance fluorescence in the statistical description of radiation emitted along two channels \textemdash{those} of forwards and sideways scattering \textemdash{comprising} the monitored output. In the simplest case of an empty cavity coupled to an external atom, we derive analytical results for the nonclassical fluctuations in the fields occupying the two channels, pursuing a mapping to the bad-cavity limit of the JC model to serve as a guide for the description of the more involved dynamics. Finally, we exemplify a conditional evolution for the composite system of a critical JC oscillator on resonance coupled to an external monitored two-level target, showing that coherent atomic oscillations of the target probe the onset of a second-order dissipative quantum phase transition in the source.
\end{abstract}


\maketitle

\section{Introduction}

The late 1980s and early 1990s witnessed the development of the formalism for describing the statistical properties of light emitted from a quantum system, driven by another nonclassical source \cite{GolubovSokolov87, TrajectoryCascCarmichael, Gardiner1993, GarninerParkinsArbitraryStat, KochanCarmichael1994}. While fundamentally interesting on its own, the theory also lends itself to the assessment of critical behaviour in non-equilibrium quantum phase transitions. In the experiment of \cite{Murch2013}, for instance, it was shown that the radiative decay rate of an atom coupled to quadrature-squeezed electromagnetic vacuum, generated by a Josephson parametric amplifier, can be reduced below its natural linewidth. This observation corroborated Gardiner's  theoretical prediction on the disparity of the rates at which the two polarization quadratures are damped when an atom interacts with a broadband squeezed vacuum \cite{Gardiner1986}. Concurrently with the latter, resonance fluorescence from a driven atom which is damped by a squeezed vacuum was studied in \cite{ResonanceFlSq1987}. Following this long path of investigation to our days, in the explicitly cascaded setup of \cite{squeezedvactomogaphy}, a weakly nonlinear system comprising a superconducting resonator coupled to an artificial atom in the dispersive regime is driven by squeezed vacuum, extending the efficient generation of squeezed states in a parametric amplifier comprising an array of Josephson junctions \cite{Squeezing2008}. Further along, an important step in characterizing the properties of squeezing via resonance fluorescence \textemdash{and} its characterstic Mollow triplet spectrum \textemdash{from} artificial atoms in circuit quantum electrodynamics (QED) was taken in \cite{ResonanceFlSqVac2016}. In what concerns now the generation of time-correlated photon pairs from sources operating at the nanoscale, inelastic tunneling of single photons has been very recently shown to produce highly bunched light in a process that can be construed as an idealized two-step cascade \cite{photonsuperbunching2019}.  

Quantum optical systems, like those mentioned above, have come to play a crucial r\^{o}le in the recent exploration of non-equilibrium phase transitions. Such dissipative quantum phase transitions rely fundamentally on the balance between output and input in a background of intense fluctuations, in contrast to their equilibrium counterparts. Shortly after the formulation of the cascaded-system theory, an experiment reported on the emission properties of two coupled cavities operating in the region of optical bistability \cite{CascadedCavities1995}. Closer to our days, the breakdown of photon blockade in zero dimensions \cite{CarmichaelPhotonBlockade}, which was experimentally demonstrated in \cite{Fink2017}, is associated with a distinct presence of quantum nonlinearity leading to a definition of a {\it strong-coupling ``thermodynamic limit''}, where fluctuations persist, while the mean-field and quantum predictions manifestly disagree. Such an out-of-equilibrium phase transition probes the paradigmatic $\sqrt{n}$ nonlinearity of the Jaynes-Cummings (JC) oscillator \cite{JCIEEE1963}, which has been revealed in a series of experiments in cavity and circuit QED (see, e.g., \cite{Wallraff2004} and \cite{Fink2008ClimbingJC}). Light-matter interaction as formulated by the driven dissipative JC model is subject to two ``thermodynamic limits'' which are fundamentally different in terms of the input-output relation they dictate. One of them is a so-called {\it weak-coupling limit}, in which quantum fluctuations reduce to an inconsequential addition (a so-called ``fuzz'') superimposed on top of the semiclassical output. The second one is a {\it strong-coupling limit}, where the system size grows together with the light-matter coupling strength and quantum fluctuations remain. Such a limit is associated with the occurrence of spontaneous dressed-state polarization and symmetry breaking on resonance \cite{Alsing_1991} and the persistence of photon blockade off-resonance \cite{CarmichaelPhotonBlockade}.  

Accessing the Fock states of a harmonic oscillator and assessing the statistical properties of the radiation emitted following excitation with a single-photon source, either coherent or incoherent, has recently revived the interest in the cascaded-systems formalism \cite{excitationSPS2016}. Subsequently, the normalized emission spectra of a two-level atom driven by the light emanating from another classically driven two-level atom were investigated in \cite{SPS22016} to be followed by a detailed analysis of a regime where ``where thermal statistics and quantum coherences coexist  and intertwine via quantum emitters,'' as demonstrated in \cite{DissipativelyMediatedCoh2017}. With regard to extended systems, a driven lattice of bidirectionally coupled cavities in the photon-blockade regime has been assigned a quasi-thermal distribution function in \cite{Flarray2019}, while a first-order dissipative quantum phase transition has recently been experimentally realized in a chain of 72 coplanar waveguide resonators \cite{DissipativeQPTcircuitQED}. Furthermore, direct correspondence between photon blockade and stationary dark-state generation has been recently explored in \cite{ChiralityBlockade2017}. Interestingly, cascaded quantum systems have also been used for a realization of a quantum information protocol in which spontaneously emitted photons from a quantum dot at a properly prepared state are collected and directed to a second quantum dot \cite{HeraldedAbsorption2017}. 

How will the crucial interplay of input and output pan out when information on the developed criticality is monitored by an external quantum system? Contextual entanglement for a laser oscillator illuminating a two-level atom was studied in \cite{Lasercohstate} in the frame of characterizing the laser output-state, while a very recent recent paper reports on a superposition of macroscopically incompatible states, localized at the maxima and the minima of the dipole potential, following a detection of the electromagnetic field \cite{maskedstates}. An investigation of the Ising quantum phase transition in a quantum magnetic field \cite{quantumIsing} came after the example of \cite{SpinCoupledIsing}, where an external spin is coupled to an Ising-type chain comprising a couple of spins; such an interaction imposes a conditional evolution on the composite system observables. In our paper, we explore the conditional evolution of a JC oscillator arising when monitoring its output by an external two-level atom. The atom polarization together with the cavity field form part of the forwards-scattering channel, which occupies a main object of our investigation. In such a cascaded setup, one aims at total absorption of the incident light, projecting the atom to its excited state with unit probability; a single photon representing the time-reversed wave packet would then be released by the atom in question in the course of spontaneous emission \cite{Leuchs_2012, Bader_2013, reversalheralded}. The single-photon wave packet must then impinge from the full $4\pi$ solid angle and have the appropriate temporal shape \cite{QEDparabolic}. 

Our discussion is organized as follows. After introducing the model in Sec. \ref{sec:modelandsetup}, based on the cascaded-systems formalism developed in \cite{TrajectoryCascCarmichael}, we isolate the internal two-level atom from the dynamics by setting its coupling strength to the cavity equal to zero in Sec. \ref{sec:cavitycoupledatom}. We provide expressions for the incoherent spectrum and the squeezing spectrum of quantum fluctuations in Secs. \ref{subsec:incoherentspectrum} and \ref{subsec:spectrumsq}, respectively, before focusing on the weak-excitation limit which preserves the state purity. We then extract an approximate formula for the second-order correlation function for forwards scattering in Sec. \ref{subsubsec:weakElimit}, which reaffirms the mapping to the bad-cavity limit. Such a correspondence gives access to a more general discussion on the second-order coherence properties in Sec. \ref{subsubsec:mappingbdl}. In the second part of our analysis, we reinstate the atom inside the cavity and assess the implications of a light-matter coupling with growing strength. In Sec. \ref{sec:adiabaticelimfield}, we remain within the framework of the bad-cavity limit permitting the adiabatic elimination of the {\it intracavity} field. In Sec. \ref{sec:phasebist}, we abandon the perturbative analysis allowed by the distinct timescales defining the the bad-cavity limit, and instead move to the strong-coupling regime, where we encounter a second-order quantum phase transition with the accompanying spontaneous symmetry breaking for the intracavity field and the associated atomic polarization. We exemplify the impact of a monitoring atom outside the cavity on the manifestation of phase bistability for a conditional evolution of the composite system in the course of single quantum trajectories. In spite of the unidirectional coupling, we find that monitoring the bistable JC oscillator has an ostensible effect on how the switching events appear; these events are correlated with disruptions in the coherent-oscillation cycles of the external atom. Some brief comments on our results and extension to future work close out the paper. 
\begin{figure}
\includegraphics[width=0.5\textwidth]{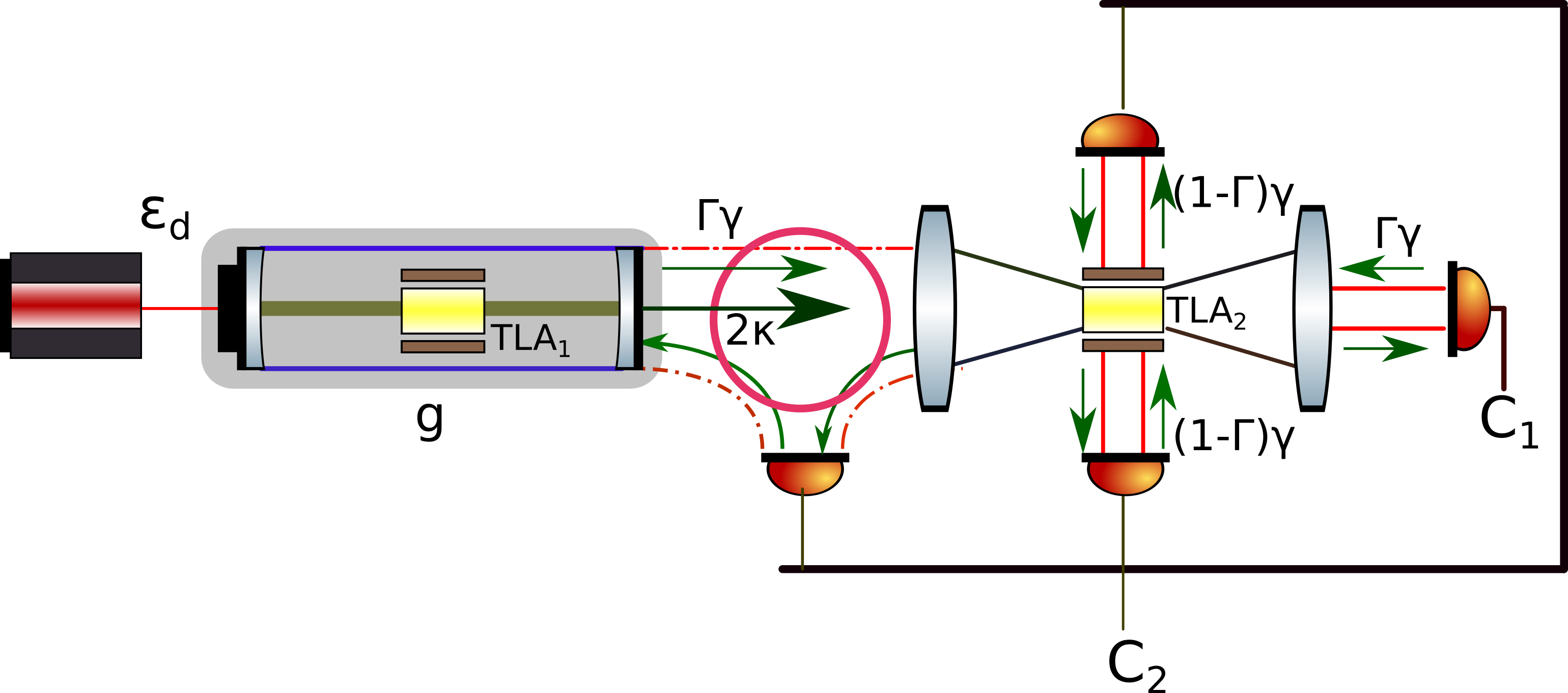}
\caption{{\it The two cascaded open quantum systems connected via the reservoir field in the vacuum state.} The modes of the vacuum field that couple to the external two-level atom (TLA$_2$) are divided between four channels. Two of them are labeled by $\Gamma \gamma$ and the other two by $(1-\Gamma)\gamma$, with $0 \leq \Gamma \leq 1$ the degree of focusing. The incident light from the driven JC oscillator occupies one channel and, superimposed with forward scattering, eventually falls on the detector corresponding to the collapse operator $C_1$. Spontaneous emission is absent for the internal two-level atom (TLA$_1$) coupled to the intracavity field with strength $g$. Backwards and sideways scattered photons are captured by the three detectors with a combined output corresponding to the collapse operator $C_2$.}
\label{fig:figureone}
\end{figure}

\section{JC oscillator coupled to a single two-level atom: the model}
\label{sec:modelandsetup}

In this work, a coherent field is driving on resonance a cavity mode coupled to a two-level atom, while the output cavity field is directed to an external two-level atom. Both atomic transitions are as well resonant with the frequency of the cavity mode. A traveling-wave reservoir connects the two subsystems unidirectionally \cite{TrajectoryCascCarmichael, Gardiner1993}. The master equation (ME) in the Markovian approximation, for the retarded density operator $\tilde{\rho}$ of the composite system {\it at the position of the external two-level atom} and in the interaction picture, reads \cite{TrajectoryCascCarmichael, DecoherenceTwostateatom, Lasercohstate} 
\begin{equation}\label{eq:ME}
\frac{d\tilde{\rho}}{dt}=\frac{1}{i\hbar}[H, \tilde{\rho}]+\sum_{k=1,2}\mathcal{L}[C_k]\tilde{\rho},
\end{equation}
where $\mathcal{L}[C_{k}]\tilde{\rho}\equiv C_{k}\tilde{\rho} C_{k}^{\dagger}-(1/2)C_{k}^{\dagger}C_{k}\tilde{\rho}-(1/2)\tilde{\rho} C_{k}^{\dagger}C_{k}$ is the standard dissipation superoperator corresponding to the collapse operator $C_{k}$ and taking as an argument the density matrix $\tilde{\rho}$. The coupled-system Hamiltonian in Eq. \eqref{eq:ME} is
\begin{equation}\label{eq:HamD}
\begin{aligned}
H= i\hbar [g(a^{\dagger}\sigma_{1-}-a\sigma_{1+})+ &\sqrt{\Gamma\kappa\gamma/4}\,(a^{\dagger}\sigma_{2-}-a\sigma_{2+})\\
& + \varepsilon_{d}(a^{\dagger}-a)],
\end{aligned}
\end{equation}
in which $g$ is the coupling strength between the cavity mode (with annihilation and creation operators $a$ and $a^{\dagger}$, respectively) and the internal atom (with polarization and inversion operators $\sigma_{1-}$ and $\sigma_{1z}$, respectively), $\varepsilon_d$ is the amplitude of the coherent field driving the cavity and $2\kappa$ is the photon loss rate due to coupling of cavity mode to a reservoir at zero temperature. The {\it total} spontaneous emission rate due to coupling of the external atom ($\sigma_{2-}$, $\sigma_{2z}$) to reservoir modes is denoted by $\gamma$, while we assume that spontaneous emission is absent for the atom inside the cavity, unless explicitly stated otherwise (e.g., in Sec. \ref{sec:adiabaticelimfield}). The fraction of the spontaneous emission rate into the solid angle subtended by the source is denoted by $\Gamma \gamma/2$, while we refer to $\Gamma$ as the {\it degree of focusing}. In such a configuration, $0 \leq \Gamma \leq 1$, with the total spontaneous emission rate being $[2\Gamma + 2(1-\Gamma)](\gamma/2)=\gamma$.  

The two collapse operators featured in Eq. \eqref{eq:ME}, reflecting the asymmetry of the channels coupling the external atom to its environment due to the presence of a degree of focusing $\Gamma$ different than unity, are (see Eq. 12 of \cite{TrajectoryCascCarmichael})
\begin{equation}
\begin{aligned}
&C_1=\sqrt{2\kappa}\,a + \sqrt{\Gamma (\gamma/2)}\, \sigma_{2-},\\
&C_2=\sqrt{[2(1-\Gamma)+\Gamma](\gamma/2)}\, \sigma_{2-}=\sqrt{(2-\Gamma)(\gamma/2)}\, \sigma_{2-},
\end{aligned}
\end{equation}
for the single forwards-scattering ($C_1$) and the collection of one backwards-scattering and two sideways-scattering channels (all lumped in $C_2$). For convenience, hereinafter we omit the designation {\it backwards-scattering} when referring to the field $C_2$ (which is, however, the dominant contribution for the limiting case $\Gamma \to 1$ considered in \citep{TrajectoryCascCarmichael}). The setup of Fig. \ref{fig:figureone}, depicting the input-output channels for an atom driven by a JC oscillator closely based on the configuration investigated in \citep{TrajectoryCascCarmichael}, is a parametrized sketch of the actual three-dimensional interactions involved (see \cite{efficientexcitation} for a discussion on the spatial overlap between the excitation pulse and the dipole moment with reference to the atomic excitation probability). The forwards-scattering channel corresponds to the superposition of two quantum fields which cannot be monitored individually without upsetting the coupling between the resonant cavity mode and the external fluorescent two-level atom. The direct coupling between the cavity field and the external two-level atom, forming part of the coherent dynamics occurs with a strength $\sqrt{\Gamma\kappa\gamma/4}$, comprised entirely of coupling rates to the reservoir fields, otherwise responsible for dissipation. We note as well that $a$ and $\sigma_{2-}$ also couple to the reservoir modes at the same spatial location \cite{TrajectoryCascCarmichael}. ME \eqref{eq:ME} is solved via exact diagonalization for the Liouvillian superoperators dictating the evolution of the composite-system density matrix. The ME is also unravelled into quantum trajectories via a quantum state diffusion algorithm (see \cite{Gisin_1992, PercivalQSD} and the correspondence with a stochastic differential equation for the continuous time evolution of conditioned heterodyne-current records, in Sec. 18.2.3 of \cite{QO2}) with adaptive stepsize. For the exact diagonalization, we use the exponential series expansion of MATLAB's {\it Quantum Optics Toolbox}, while for the generation of individual realizations we rely on an open-source library in C\texttt{++} detailed in \cite{qsdreference}. 

To gain an understanding of where the unidirectional coupling could lead to, we begin by looking at the mean-field equations. The semiclassical equations for $\tilde{\alpha}\equiv\braket{a}$, $\tilde{\beta}_1\equiv\braket{\sigma_{1-}}$, $\zeta_1\equiv\braket{\sigma_{1z}}$, $\tilde{\beta}_2\equiv\braket{\sigma_{2-}}$, $\zeta_2\equiv\braket{\sigma_{2z}}$, derived from the ME \eqref{eq:ME} after factorizing the coupled moments in the equations of motion, read
\begin{subequations}\label{eq:neocl2}
\begin{align}
&\frac{d\tilde{\alpha}}{dt}=-\kappa\tilde{\alpha} +g \tilde{\beta}_1 +\varepsilon_d,   \label{eq:neocl2a}\\
&\frac{d\tilde{\beta}_1}{dt}= g \tilde{\alpha}\zeta_1, \label{eq:neocl2b}\\
&\frac{d\zeta_1}{dt}=-2g(\tilde{\alpha}^{*}\tilde{\beta}_1+\tilde{\alpha}\tilde{\beta}_1^{*}), \label{eq:neocl2c}\\
&\frac{d\tilde{\beta}_2}{dt}=-(\gamma/2)\, \tilde{\beta}_2 + \sqrt{\kappa\gamma\Gamma}\,\tilde{\alpha}\zeta_2, \label{eq:neocl2d} \\
&\frac{d\zeta_2}{dt}=-\gamma(\zeta_2+1) - 2 \sqrt{\kappa\gamma\Gamma}(\tilde{\alpha}^{*}\tilde{\beta}_2+\tilde{\alpha} \tilde{\beta}_2^{*}). \label{eq:neocl2e}
\end{align}
\end{subequations}
Since spontaneous emission is absent for the atom lying inside the cavity, Eqs. \eqref{eq:neocl2b} and \eqref{eq:neocl2c} preserve the length of the pseudo-spin for the internal two-level atom interacting with the resonant cavity mode, yielding \cite{Alsing_1991, CarmichaelPhotonBlockade}
\begin{equation}\label{eq:conservation}
4|\tilde{\beta}_1|^2 + \zeta_1^2=1.
\end{equation} 

Eqs. \eqref{eq:neocl2a} - \eqref{eq:neocl2c} predict the appearance of spontaneous dressed-state polarization for the JC ``molecule'' when $\varepsilon_d \geq g/2$, producing states which become attractors in the presence of quantum fluctuations \cite{Alsing_1991}. We also note that {\it on the mean-field level}, the equations of motion for the atomic averages are the same as those of free-space resonance fluorescence, where the atom is driven by a coherent field with complex amplitude $\tilde{\alpha}$. This is a consequence of the unidirectional coupling. In the steady state, we then find
\begin{equation}\label{eq:steadystateMF}
\tilde{\beta}_{2, \,{\rm ss}}=-\frac{1}{\sqrt{2}} \frac{Y}{1+|Y|^2}, \quad \quad \zeta_{2, \,{\rm ss}}=-\frac{1}{1+|Y|^2},
\end{equation}
with the dimensionless drive amplitude defined as $Y \equiv 2\sqrt{2\kappa\Gamma/\gamma}\,\tilde{\alpha}_{\rm ss}$. Having now introduced the model we will be working with, we proceed to a significant simplification by considering an empty cavity driven by coherent light, producing an output field which is directed to the external two-level atom. 

\section{Coherently driven empty cavity coupled to an external two-level atom} 
\label{sec:cavitycoupledatom}

Let us now consider the case where $g=0$. For brevity we drop the subscript of the atomic operators, reserving $\sigma$ for the external atom only (since the internal atom has no longer any influence on the dynamics). In this section, we draw motivation by the analysis of the same system considered in \cite{TrajectoryCascCarmichael}, which we briefly summarize in the following paragraphs.

In the interaction picture, the non-Hermitian Hamiltonian governing the evolution of the (un-normalized) conditional wave-function $\ket{\psi_c(t)}$, as $(d/dt)(\ket{\psi_c(t)})=[1/(i\hbar)] \mathcal{H}\ket{\psi_c(t)}$ has the form
\begin{equation}\label{eq:nonHerm}
\mathcal{H}=i\hbar[\varepsilon_d(a^{\dagger}-a)-\kappa a^{\dagger}a - (\gamma/2)\sigma_{+}\sigma_{-}-\sqrt{2\Gamma\kappa (\gamma/2)}\,a \sigma_{+}].
\end{equation}
Assuming an initial vacuum state for the cavity mode and given that the term $a^{\dagger}\sigma_{-}$ is absent from the Hamiltonian of Eq. \eqref{eq:nonHerm}, the conditional wavefunction can be written in the factorized form $\ket{\psi_c(t)}=\ket{\alpha(t)}\ket{A_c(t)}$, where $\alpha(t)$ is a coherent-state amplitude and $\ket{A_c(t)}$ is the state of the atom. When the field is in a coherent state, the atom does not entangle with its driving field \cite{DecoherenceTwostateatom}. We find that the amplitude $\alpha(t)$ is given by the expression $\alpha(t)=(\varepsilon_d/\kappa)(1-e^{-\kappa t})$ and, upon reaching steady state, we obtain $\alpha_{\rm ss}=\varepsilon_d/\kappa$. Then, in the long-time limit, the wavefunction for the (external) atom alone obeys
\begin{equation}\label{eq:nonHermAtom}
\frac{d}{dt}\ket{A_c(t)}=-\left(\frac{\gamma}{2}\sigma_{+}\sigma_{-} + \varepsilon_d \sqrt{\frac{\Gamma\gamma}{\kappa}} \sigma_{+}\right)\ket{A_c(t)},
\end{equation}
with collapse operators $C_1=\sqrt{2\kappa}(\varepsilon_d/\kappa)+\sqrt{\Gamma (\gamma/2)}\sigma_{-}$ and $C_2=\sqrt{(2-\Gamma)(\gamma/2)}\, \sigma_{-}$.

Now, in \cite{TrajectoryCascCarmichael} we also read that this evolution is equivalent to placing the atom {\it inside} a driven cavity in the bad-cavity limit, satisfying a ME of the open driven JC oscillator on resonance (in the interaction picture)
\begin{equation}
\begin{aligned}
\frac{d \tilde{\rho}}{dt}&=\overline{g}[a^{\dagger}\sigma_{-}-a\sigma_{+}, \tilde{\rho}]+ \varepsilon_d [a^{\dagger}-a, \tilde{\rho}]\\
&+\kappa \left(2a \tilde{\rho} a^{\dagger} - a^{\dagger}a \tilde{\rho} - \tilde{\rho} a^{\dagger}a\right)\\
&+\frac{\overline{\gamma}_{\rm s}}{2}\left(2\sigma_{-}\tilde{\rho} \sigma_{+}-\sigma_{+}\sigma_{-}\tilde{\rho} - \tilde{\rho} \sigma_{+}\sigma_{-}\right),
\end{aligned}
\end{equation}
with $\overline{g} \equiv \sqrt{\kappa \Gamma \gamma/2}$ and $\overline{\gamma}_{\rm s}=\sqrt{(2-2\Gamma)(\gamma/2)}$, adopting the notation of \cite{TrajectoryCascCarmichael}. In the bad-cavity limit, with the adiabatic elimination of the cavity field being justified when $\kappa \gg (\gamma/2, g)$, the {\it intracavity} field operator $a$ is then identifiable with the forwards-scattering field operator (in units of the square root of photon flux) $C_1 \equiv \sqrt{2\kappa}(\varepsilon_d/\kappa) + \sqrt{\Gamma(\gamma/2)} \sigma_{-}$, the statistics of which we wish to determine. This mapping  would also lead to an enhanced emission rate of the form $(1+2C) \overline{\gamma}_{\rm s}$, with $C\equiv \overline{g}^2/(\kappa\overline{\gamma}_{\rm s})=\Gamma/[2(1-\Gamma)]$, whence $(1+2C) \overline{\gamma}_{\rm s}=[1+\Gamma/(1-\Gamma)]2(1-\Gamma)(\gamma/2)=\gamma$. In this correspondence, $\Gamma=2C/(1+2C)$ is the proportion of the atomic reradiation inside the cavity seen in transmission \cite{RiceCarmichaelNonClassical}. We will discuss this mapping in more detail in Sec. \ref{subsubsec:mappingbdl}.

For the moment, we return to the solution of ME \eqref{eq:ME} in the case where the atom inside the cavity is explicitly not involved.  As in Sec. V of \cite{DecoherenceTwostateatom} and Sec. IIB of \cite{KochanCarmichael1994}, we propose the {\it ansatz}
\begin{equation}\label{eq:asnatzcoh}
\tilde{\rho}(t)=|\alpha(t)\rangle \langle\alpha(t)| \otimes \rho_A(t),
\end{equation}
leading to a reduced ME for the external atom alone, 
\begin{equation}\label{eq:MEatom}
\frac{d{\rho}_A}{dt}=\frac{1}{i\hbar}[H_{\rm eff}, \rho_A] + \mathcal{L}[C_A]\rho_A,
\end{equation}
with an effective Hamiltonian
\begin{equation}\label{eq:Heff}
\begin{aligned}
H_{\rm eff}&=i\hbar\sqrt{2\kappa \Gamma (\gamma/2)}[\alpha^{*}(t)\sigma_{-}-\alpha(t)\sigma_{+}]\\
&=i\hbar\sqrt{\kappa \Gamma \gamma}[\alpha^{*}(t)\sigma_{-}-\alpha(t)\sigma_{+}],
\end{aligned}
\end{equation}
and a single collapse operator
\begin{equation}\label{eq:CA}
C_A=\sqrt{\gamma}\,\sigma_{-}.
\end{equation}
The coherent-state amplitude evolves again as $\alpha(t)=(\varepsilon_d/\kappa)(1-e^{-\kappa t})$ for a time-independent coherent drive, relaxing to the steady-state value $\alpha_{\rm ss}=\varepsilon_d/\kappa$. Since the cavity is in a coherent state, the neoclassical equations \eqref{eq:neocl2d} and \eqref{eq:neocl2e} are identical to the Heisenberg equations of motion with a steady-state solution given by Eq. \eqref{eq:steadystateMF}, where $Y=2 \sqrt{2}\,\varepsilon_d\sqrt{\Gamma/(\kappa\gamma)}$.

\subsection{Incoherent spectrum of fluctuations for the two channels}
\label{subsec:incoherentspectrum}

We will now carry on with the ME produced for the external atom alone which, after the field amplitude has relaxed to its final value, can be written in the standard form of free-space resonance fluorescence, as
\begin{equation}\label{eq:MEatomalone}
\frac{d{\rho}_A}{dt}=\frac{1}{i\hbar}[H_{\rm eff, (t\gg \kappa^{-1})}, \rho_A] + \mathcal{L}(C_A)\rho_A,
\end{equation}
in which the effective Hamiltonian has relaxed to
\begin{equation}\label{eq:HamSteadyState}
\begin{aligned}
H_{\rm eff, (t \gg \kappa^{-1})}&=(1/2)\hbar \omega_A \sigma_z\\
& + i\hbar \varepsilon_d\sqrt{\Gamma \gamma/\kappa}(\sigma_{-}e^{i\omega_A t}-\sigma_{+}e^{-i\omega_A t}),
\end{aligned}
\end{equation}
where $\omega_A$ is the atomic frequency (coinciding with the frequency of the drive and the resonance frequency of the intracavity mode). In the Appendix, we derive the correlation functions needed to calculate the incoherent spectrum of the forwards-emitted field from the steady-state first-order correlation function \cite{QO1}. These are the same as in ordinary resonance fluorescence since the fluctuations $\Delta C_{1,2}$ and $\Delta C_{1,2}^{\dagger}$ (where $\Delta C_{1,2} \equiv C_{1,2}-\braket{C_{1,2}}_{\rm ss}$) are proportional to $\Delta\sigma_{-}$ and $\Delta \sigma_{+}$, respectively, and the quantum regression formula relies on the Bloch equations (in which appropriately modified coefficients feature). For all the steady-state averages, $\braket{\cdot}_{\rm ss}$, the limit $t\to \infty$ has already been attained. Therefore, one only requires the Hamiltonian of Eq. \eqref{eq:HamSteadyState} when assessing the coherence properties of the source field radiated by the atom outside the cavity as a {\it stationary} process. 

Adopting the scaling of \cite{RiceCarmichaelNonClassical} and following the standard procedure (see, e.g., Sec. 2.3.4 of \cite{QO1}), we write the incoherent optical spectrum of the forwards and sideways scattered fields as the Fourier transform of the first-order fluctuation correlation function for the slowly varying operators $\tilde{C}_1, \tilde{C}_2$, at a scaled angular frequency displaced by the atomic resonance, as
\begin{equation}
\begin{aligned}
S_{\rm C_{1,2},\,inc}(\overline{\omega})&=\frac{1}{2\pi}(\braket{\Delta\tilde{C}_{1,2}^{\dagger}\Delta \tilde{C}_{1,2}}_{\rm ss})^{-1}\\
&\times \int_{-\infty}^{\infty}d\overline{\tau} e^{i(\overline{\omega}-\overline{\omega}_A)\overline{\tau}}\braket{\Delta\tilde{C}^{\dagger}_{1,2}(0)\Delta \tilde{C}_{1,2}(\overline{\tau})}_{\rm ss}\\
=&\frac{1}{\pi} \left(\frac{1}{2}\frac{Y^4}{(1+Y^2)^2}\right)^{-1}\\
&\times {\rm Re}\left(\int_{0}^{\infty}d\overline{\tau} e^{i(\overline{\omega}-\overline{\omega}_A)\overline{\tau}}\braket{\Delta\tilde{\sigma}_{+}(0)\Delta \tilde{\sigma}_{-}(\overline{\tau})}_{\rm ss} \right),
\end{aligned}
\end{equation}
where the (slowly varying) operators $\tilde{C}_{1,2}(t)$ and $\tilde{C}^{\dagger}_{1,2}(t)$ are defined in a frame rotating with $\omega_A$. The incoherent spectrum evaluates to (see Eq. 1 of \cite{RiceCarmichaelNonClassical} and Eq. 22 of \cite{SPS22016})
\begin{equation}\label{eq:incohspectrfinal}
\begin{aligned}
&S_{\rm C_{1,2},\,inc}(\overline{\omega})=\frac{1}{2\pi}\Bigg\{\left(\frac{Y^2}{1+Y^2}\right)^{-1} \frac{1}{1+(\overline{\omega}-\overline{\omega}_A)^2}\\
&-\left[1/Y^2-1+(1/Y^2-5)\frac{1}{2\overline{\delta}}\right]\frac{3/4-\overline{\delta}/2}{(3/2-\overline{\delta})^2+(\overline{\omega}-\overline{\omega}_A)^2}\\
&- \left[1/Y^2-1-(1/Y^2-5)\frac{1}{2\overline{\delta}}\right]\frac{3/4+\overline{\delta}/2}{(3/2+\overline{\delta})^2+(\overline{\omega}-\overline{\omega}_A)^2}\Bigg\},
\end{aligned}
\end{equation}
where $\overline{\tau} \equiv \gamma \tau/2$, $\overline{\omega} \equiv 2\omega/\gamma$, $\overline{\omega}_A \equiv 2\omega_A/\gamma$ and $\overline{\delta}\equiv 2\delta/\gamma$, with $\delta=(\gamma/4)\sqrt{1-8Y^2}$. The above expression is normalized (to unit area) with respect to the dimensionless angular frequency $\overline{\omega}$. At the exceptional point, $\delta=0$ (see the Appendix), the incoherent spectrum is given by the expression [see Eq. (5) of \cite{RiceCarmichaelNonClassical}]
\begin{equation}\label{eq:incohcr}
\begin{aligned}
S_{\rm C_{1,2},\,inc,\,cr}(\overline{\omega})&=\frac{9}{2\pi}\Bigg\{\frac{1}{1+(\overline{\omega}-\overline{\omega}_A)^2}\\
& - \frac{3+(\overline{\omega}-\overline{\omega}_A)^2}{[(3/2)^2+(\overline{\omega}-\overline{\omega}_A)^2]^2}\Bigg\},
\end{aligned}
\end{equation}
yielding a narrower distribution than the free-space Lorentzian spectrum, before the Rabi doublet emerges.

\subsection{Squeezing of quantum fluctuations and the spectrum of squeezing}
\label{subsec:spectrumsq}

The incoherent spectrum of the quantum fields occupying the two channels, corresponding to the operators $C_1, C_2$, is intimately tied to the spectrum of squeezing which, unlike the former, can assume negative values. Essentially, the spectrum of squeezing for both channels assumes the same form as in ordinary resonance fluorescence, since the source operators $\Delta \sigma_{\pm}$ obey the same optical Bloch equations as we have already pointed out (for a discussion on the self-homodyning of squeezed florescence and its contribution to antibunching see, e.g., the form Eq. 37 and the ensuing discussion in \cite{RiceCarmichaelIEEE}, and compare to Sec. 2.3.6 of \cite{QO1}). Defining the field quadratures $\Delta \tilde{X}_{1,2} \equiv \tilde{X}_{1,2}-\braket{\tilde{X}_{1,2}}_{\rm ss}$ with $\sqrt{2\kappa} \tilde{X}_{1}\equiv (1/2)(\tilde{C}_1+\tilde{C}_1^{\dagger})$ and $\sqrt{2\kappa} \tilde{X}_2 \equiv -i(1/2)(\tilde{C}_1-\tilde{C}_1^{\dagger})$, we calculate the normally ordered quadrature variances in the steady state (see also Eq. 32 of \cite{RiceCarmichaelIEEE}), 
\begin{equation}\label{eq:variancesnormord}
\begin{aligned}
&\braket{:(\Delta \tilde{X}_{1,2})^2:}_{\rm ss}\\
&=\frac{1}{4}[2\braket{\Delta \tilde{C}_1^{\dagger} \Delta \tilde{C}_1}_{\rm ss}\pm \braket{(\Delta \tilde{C}_1)^2}_{\rm ss} \pm \braket{(\Delta \tilde{C}_1^{\dagger})^2}_{\rm ss}]\\
&=\frac{1}{4} \left(\frac{\Gamma \gamma}{4\kappa}\right)\left[1+\braket{\sigma_z}_{\rm ss} - (2 \pm 2) \braket{\tilde{\sigma}_{+}}^2_{\rm ss} \right]\\
&=\frac{1}{4} \left(\frac{\Gamma \gamma}{4\kappa}\right)\frac{Y^2(Y^2 \mp 1)}{(1+Y^2)^2}.
\end{aligned}
\end{equation}
Squeezing of the steady-state quantum fluctuations occurs only for the field quadrature $\tilde{X}_1$, which is in phase with the steady-state polarization $\braket{\tilde{\sigma}_{-}}_{\rm ss}$, for $Y<1$; this variance is an explicit function of the degree of focusing.

The spectrum of squeezing for the outward-field quadrature, as measured via a homodyne detection scheme employing a local oscillator with phase $\theta$, is \cite{RiceCarmichaelNonClassical} 
\begin{equation}\label{eq:spectrsq}
\begin{aligned}
&S_{\rm C_1,\,sq}(\overline{\omega}, \theta)=\frac{8\eta}{\pi}\int_{0}^{\infty}d\overline{\tau}\cos(\overline{\omega}\,\overline{\tau})\\
& \times {\rm Re}\left(\braket{\Delta\tilde{C}^{\dagger}_{1}(0)\Delta \tilde{C}_{1}(\overline{\tau})}_{\rm ss} + e^{2i\theta} \braket{\Delta\tilde{C}^{\dagger}_{1}(0)\Delta \tilde{C}^{\dagger}_{1}(\overline{\tau})}_{\rm ss}\right),
\end{aligned}
\end{equation}
where $\eta$ stands for the product of the collection and detection coefficients. Once more, we need to sequester formulas of ordinary resonance fluorescence from the Appendix. For $\theta=0$, the quantity in the integral of Eq. \eqref{eq:spectrsq} is the normally ordered correlation function of the in-phase quadrature $X_1$, which, as we anticipate when $\tau=0$, is explicitly negative for $Y \ll 1$ according to the calculation below:
\begin{equation}\label{eq:corrXsq}
\begin{aligned}
&\braket{:\Delta \tilde{X}_1(0)\Delta \tilde{X}_1(\tau):}_{\rm ss}\\
&=\frac{1}{2}\left(\frac{\Gamma\gamma}{4\kappa}\right)[\braket{\Delta\tilde{\sigma}_{+}(0)\Delta\tilde{\sigma}_{+}(\tau)}_{\rm ss} + \braket{\Delta\tilde{\sigma}_{+}(0)\Delta\tilde{\sigma}_{-}(\tau)}_{\rm ss}]\\
&=-\frac{1}{4}\left(\frac{\Gamma\gamma}{4\kappa}\right)\frac{Y^2}{(1+Y^2)^2}\\
&\times\Bigg\{\left[1-Y^2 + \left(\frac{\gamma}{4\delta}\right)(1-5Y^2)\right]e^{-(3\gamma/4-\delta)\tau}\\
&+\left[1-Y^2 - \left(\frac{\gamma}{4\delta}\right)(1-5Y^2)\right]e^{-(3\gamma/4+\delta)\tau}\Bigg\}.
\end{aligned}
\end{equation}

\begin{figure*}
\includegraphics[width=\textwidth]{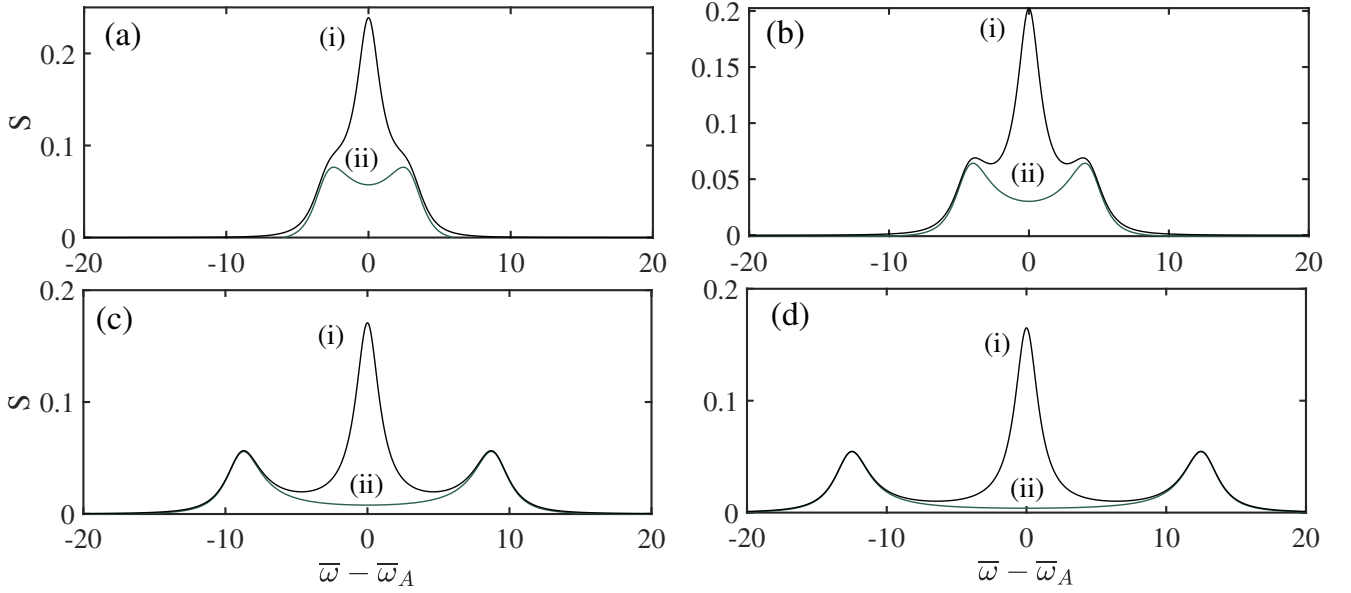}
\caption{{\it Incoherent scattering and squeezing of fluctuations in the forwards-scattering channel for increasing degree of focusing.} Incoherent spectra $S_{\rm C_{1},\,inc}(\overline{\omega})$ [curves (i)] obtained from Eq. \eqref{eq:incohspectrfinal}, showing the development of the Mollow triplet, and normalized squeezing spectra $\overline{S}_{\rm C_1,\,sq}(\overline{\omega}-\overline{\omega}_A)$ [curves (ii)] obtained from Eq. \eqref{eq:sqztheta0eval} (centered at $\overline{\omega}_A$) of the forwards-propagating field for an increasing degree of focusing: $\Gamma=0.05, 0.1, 0.4, 0.8$ in frames {\bf (a)}-{\bf (d)}, respectively. The remaining parameters read: $\kappa/\gamma=200$ and $\varepsilon_d/\gamma=50$.}
\label{fig:figuretwo}
\end{figure*}

From this function, one computes the squeezing spectrum for the field quadrature of forwards-scattered light which is in phase with the induced atomic polarization as
\begin{equation}\label{eq:sqztheta0eval}
\begin{aligned}
&S_{\rm C_1,\,sq}(\overline{\omega}, 0)=-\frac{2\eta}{\pi}\left(\frac{\Gamma\gamma}{2}\right) \frac{Y^2}{(1+Y^2)^2}\\
&\times\Bigg\{\left[1-Y^2 + \left(\frac{1}{2\overline{\delta}}\right)(1-5Y^2)\right]\frac{3/2-\overline{\delta}}{(3/2-\overline{\delta})^2+\overline{\omega}^2}\\
&+\left[1-Y^2 - \left(\frac{1}{2\overline{\delta}}\right)(1-5Y^2)\right]\frac{3/2+\overline{\delta}}{(3/2+\overline{\delta})^2+\overline{\omega}^2}\Bigg\}.
\end{aligned}
\end{equation}
On the other hand, for the fluctuations in quadrature to $\braket{\tilde{\sigma}_{-}}_{\rm ss}$ ($\theta=\pi/2$), we obtain a standard Lorentzian with natural linewidth,
\begin{equation}\label{eq:sqinquadr}
\begin{aligned}
&S_{\rm C_1,\,sq}(\overline{\omega}, \pi/2)=(2\kappa)\frac{8\eta}{\pi}\\
& \times \int_{0}^{\infty}d\overline{\tau}\cos(\overline{\omega}\,\overline{\tau})\braket{:\Delta \tilde{X}_2(0)\Delta \tilde{X}_2(\tau):}_{\rm ss}\\
&=\frac{8\eta}{\pi}\left(\frac{\Gamma\gamma}{2}\right) \int_{0}^{\infty}d\overline{\tau}\cos(\overline{\omega}\,\overline{\tau})\\
&\times \left(\braket{\Delta\tilde{\sigma}_{+}(0)\Delta \tilde{\sigma}_{-}(\overline{\tau})}_{\rm ss} - \braket{\Delta\tilde{\sigma}_{+}(0)\Delta \tilde{\sigma}_{+}(\overline{\tau})}_{\rm ss} \right)\\
&=\frac{4\eta}{\pi}\left(\frac{\Gamma\gamma}{2}\right)\frac{Y^2}{1+Y^2}\frac{1}{1+\overline{\omega}^2}.
\end{aligned}
\end{equation}
Hence, we recover the general result linking the incoherent spectrum to the spectrum of squeezing \cite{RiceCarmichaelNonClassical}
\begin{equation}\label{eq:linkincsq}
\begin{aligned}
S_{\rm C_{1},\,inc}(\overline{\omega}+\overline{\omega}_A)&=(16\eta \braket{\Delta\tilde{C}_{1}^{\dagger}\Delta\tilde{C}_1}_{\rm ss})^{-1}\\
& \times[S_{\rm C_1,\,sq}(\overline{\omega}, \theta)+S_{\rm C_1,\,sq}(\overline{\omega}, \theta+\pi/2)],
\end{aligned}
\end{equation}
applying in our case for $\theta=0$ (see also Fig. 2 of \cite{SPS22016} for an explicit formation of the incoherent spectrum as a balance of Lorentzians with positive and negative weights). For weak excitation strengths, $Y^2 \ll 1$, the spectrum $S_{\rm C_1,\,sq}(\overline{\omega}, 0)$ takes negative values due to squeezing of fluctuations in phase with the mean induced polarization; this lies at the root of the squared Lorentzian profile whose origins date back to Mollow as reported in 1969 [see \cite{RiceCarmichaelNonClassical} as well as Eq. (4.21) of \cite{Mollow1969} and discussion below]. The incoherent spectrum of Eq. \eqref{eq:incohspectrfinal} and the normalized squeezing spectra as given by
\begin{equation}\label{eq:normSq}
\overline{S}_{\rm C_1,\,sq}(\overline{\omega}) \equiv \left(16\eta \braket{\Delta \tilde{C}_1^{\dagger}\Delta \tilde{C}_1}_{\rm ss}\right)^{-1} S_{\rm C_1,\,sq}(\overline{\omega})
\end{equation}
for the forwards emission are depicted in Fig. \ref{fig:figuretwo}. In this figure, where $Y>1/(2\sqrt{2})$ for all frames, we witness the development of the characteristic Mollow triplet for an increasing degree of focusing, a sign of dominant incoherent scattering. The Rabi sidebands of Figs. \ref{fig:figuretwo}(c) and \ref{fig:figuretwo}(d) are perfectly captured by the spectrum of squeezing for strong focusing, as predicted by the dominant contribution of the second term in the sum of Eq. \eqref{eq:linkincsq} to the incoherent spectrum, for large values of $Y$.

\subsection{Second-order coherence for the two channels}

To get a deeper insight for the statistics of the forwards and sideways-scattered light we now consider the second-order correlators. The second-order correlation function for the forwards-scattering field, corresponding to the operator $C_1$, is: 
\begin{equation}\label{eq:g2fdefinition}
\begin{aligned}
g^{(2)}_{C_1}(\tau)&=\frac{\braket{\tilde{C}_1^{\dagger}(0)\tilde{C}_1^{\dagger}(\tau)\tilde{C}_1(\tau)\tilde{C}_1(0)}_{\rm ss}}{\braket{\tilde{C_1}^{\dagger}\tilde{C}_1}_{\rm ss}^2} \\
&=\frac{{\rm tr}\{\tilde{C}_1^{\dagger}(0)\tilde{C}_1(0) e^{\tilde{\mathcal{L}}\tau}[\tilde{C}_1(0)\tilde{\rho}_{\rm ss}\tilde{C}_1^{\dagger}(0)]\}}{\braket{\tilde{C}_1^{\dagger}\tilde{C}_1}_{\rm ss}^2}\\
&=\frac{\braket{(\tilde{C}^{\dagger}_1\tilde{C}_1)(\tau)}_{\tilde{\rho}(0)=\tilde{\rho}^{\prime}_{\rm ss}}}{\braket{\tilde{C}_1^{\dagger}\tilde{C}_1}_{\rm ss}},
\end{aligned}
\end{equation}  
where in passing from the first to the second line we have once more employed the quantum regression formula. The (normalized) initial state of the atomic system $\tilde{\rho}(0)=\tilde{\rho}_{\rm ss}^{\prime}$, for which the above averages are evaluated, is given by
\begin{equation}\label{eq:reducedstate}
\tilde{\rho}_{\rm ss}^{\prime} \equiv \frac{\tilde{C}_1 \tilde{\rho}_{\rm ss} \tilde{C}_{1}^{\dagger}}{{\rm tr}(\tilde{C}_1 \tilde{\rho}_{\rm ss} \tilde{C}_{1}^{\dagger})}.
\end{equation}
We note here that the steady-state mean photon flux in the forwards direction, featuring in the above expression, is given by the expression 
\begin{equation}\label{eq:flux}
\begin{aligned}
\braket{\tilde{C}^{\dagger}_1 \tilde{C}_1}_{\rm ss}&=(2\kappa)\frac{1}{8\Gamma}\frac{\gamma}{\kappa}\frac{Y^2}{1+Y^2}[(1-\Gamma)^2+Y^2]\\
&=\frac{\gamma}{4\Gamma}\frac{Y^2}{1+Y^2}[(1-\Gamma)^2+Y^2].
\end{aligned}
\end{equation}
For weak driving, $Y^2 \ll 1$, this quantity reduces to
\begin{equation}\label{eq:fluxweak}
\braket{\tilde{C}^{\dagger}_1 \tilde{C}_1}_{\rm ss} \approx (2\kappa)\left(\frac{\varepsilon_d}{\kappa}\right)^2 (1-\Gamma)^2,
\end{equation}
provided that $\Gamma$ is not too close to unity.

\subsubsection{The weak-excitation limit}
\label{subsubsec:weakElimit}

The calculation is significantly simplified in the low-excitation limit. We follow closely the treatment in Sec. 13.2.3 of \cite{QO2} regarding the statistics of a weak intracavity field in the bad-cavity limit. Within the Hilbert space of the external atom, the density matrix in the steady state, following Eq. \eqref{eq:steadystateMF} with (the real) $Y=2\sqrt{2} \varepsilon_d\sqrt{\Gamma/(\kappa\gamma)}$, is
\begin{equation}\label{eq:Steadystateatom}
\begin{aligned}
(\tilde{\rho}_A)_{\rm ss}&=\frac{1}{2}\frac{2+Y^2}{1+Y^2}|1 \rangle \langle 1| + \frac{1}{2}\frac{Y^2}{1+Y^2}|2 \rangle \langle 2|\\
&-\frac{1}{\sqrt{2}}\frac{Y}{1+Y^2}(|1 \rangle \langle 2|+|2 \rangle \langle 1|).
\end{aligned}
\end{equation}
In the limit $Y \ll 1$, the steady state may then be approximated by a pure state as
\begin{equation}\label{eq:purestateappr}
(\tilde{\rho}_A)_{\rm ss} \approx |\tilde{A}_{\rm ss}\rangle \langle \tilde{A}_{\rm ss}|,
\end{equation}
with
\begin{equation}\label{eq:state}
\ket{\tilde{A}_{\rm ss}}=\ket{1}-\frac{1}{\sqrt{2}}Y\ket{2}.
\end{equation}
The reduced state given by Eq. \eqref{eq:reducedstate}, prepared under the condition that a photodetection occurs in the forward direction at $\tau=0$, is then also approximately pure and can be written in the factorized form
\begin{equation}\label{eq:redstateappr}
(\tilde{\rho}_A)^{\prime}_{\rm ss} \approx |\tilde{A}^{\prime}_{\rm ss}\rangle \langle \tilde{A}^{\prime}_{\rm ss}|,
\end{equation}
with
\begin{equation}\label{eq:redstatefull}
\begin{aligned}
&\ket{\tilde{A}^{\prime}_{\rm ss}}=\frac{\tilde{C}_1 \ket{\tilde{A}_{\rm ss}}}{\sqrt{\braket{\tilde{A}_{\rm ss}|\tilde{C}_1^{\dagger}\tilde{C}_1|\tilde{A}_{\rm ss}}}}=\\
&\frac{[\varepsilon_d/\kappa+\sqrt{\Gamma \gamma/(4\kappa)}\,\tilde{\sigma}_{-}]\ket{\tilde{A}_{\rm ss}}}{\sqrt{\braket{\tilde{A}_{\rm ss}| [\varepsilon_d/\kappa+\sqrt{\Gamma \gamma/(4\kappa)}\,\tilde{\sigma}_{+}][\varepsilon_d/\kappa+\sqrt{\Gamma \gamma/(4\kappa)}\,\tilde{\sigma}_{-}] |\tilde{A}_{\rm ss}}}},
\end{aligned}
\end{equation}
For the un-normalized state in the numerator of Eq. \eqref{eq:redstatefull} we write
\begin{equation*}
\begin{aligned}
[\varepsilon_d/\kappa+\sqrt{\Gamma \gamma/(4\kappa)}\,\tilde{\sigma}_{-}]\ket{\tilde{A}_{\rm ss}}=\left(\varepsilon_d/\kappa+\sqrt{\Gamma \gamma/(4\kappa)}\,\tilde{\sigma}_{-}\right)[\ket{1}-(Y/\sqrt{2})\ket{2}]\\
=\left(\varepsilon_d/\kappa-\sqrt{\Gamma \gamma/(8\kappa)}\right)\ket{1}-(\varepsilon_d/\kappa)(Y/\sqrt{2})\ket{2}. 
\end{aligned}
\end{equation*}
To dominant order in the driving-field amplitude, the state norm, equal to the square of the denominator in Eq. \eqref{eq:redstatefull}, is
\begin{equation}\label{eq:dominantmod}
\begin{aligned}
\frac{1}{2\kappa} {\rm tr}(\tilde{C}_1 \tilde{\rho}_{\rm ss} \tilde{C}_1^{\dagger})& \approx \left(\frac{\varepsilon_d}{\kappa}-2\frac{\varepsilon_d}{\kappa}\sqrt{\frac{\Gamma\gamma}{4\kappa}}\sqrt{\frac{\kappa\Gamma}{\gamma}}\right)^2\\
&=(\varepsilon_d/\kappa)^2 (1-\Gamma)^2,
\end{aligned}
\end{equation}
an expression we have already met in Eq. \eqref{eq:fluxweak}. Bringing the different pieces together, we write the reduced state in the same order of magnitude with respect to $Y$ as 
\begin{equation}\label{eq:redstateapp}
\ket{\tilde{A}^{\prime}_{\rm ss}} \approx \ket{1}-\frac{1}{1-\Gamma}\frac{Y}{\sqrt{2}} \ket{2}=\ket{1}-2\frac{\varepsilon_d}{\kappa}\sqrt{\frac{\kappa\Gamma}{\gamma(1-\Gamma)^2}}\, \ket{2}.
\end{equation}
The relaxation of the conditional state $\ket{\tilde{A}^{\prime}_{\rm ss}}$ to the steady state $\ket{\tilde{A}_{\rm ss}}$ occurs via the action of the propagator $e^{\tilde{\mathcal{L}}_A \tau}$, where the Liouvillian super-operator $\tilde{\mathcal{L}}_A$ is defined through the ME of Eq. \eqref{eq:MEatom}:
\begin{equation}\label{eq:Liouvillian}
\begin{aligned}
\tilde{\mathcal{L}}_A &\equiv -\varepsilon_d \sqrt{\Gamma\gamma/\kappa}\, [\sigma_{+}-\sigma_{-}, \cdot]\\
& + \frac{\gamma}{2}(2\sigma_{-}\cdot \sigma_{+}-\sigma_{+}\sigma_{-}\cdot-\cdot \sigma_{+}\sigma_{-})\\
&\approx -\varepsilon_d \sqrt{\Gamma\gamma/\kappa}\,[\sigma_{+}, \cdot]-\frac{\gamma}{2}\{\sigma_{+}\sigma_{-}, \cdot\}.
\end{aligned}
\end{equation}
\begin{figure}
\begin{center}
\includegraphics[width=0.5\textwidth]{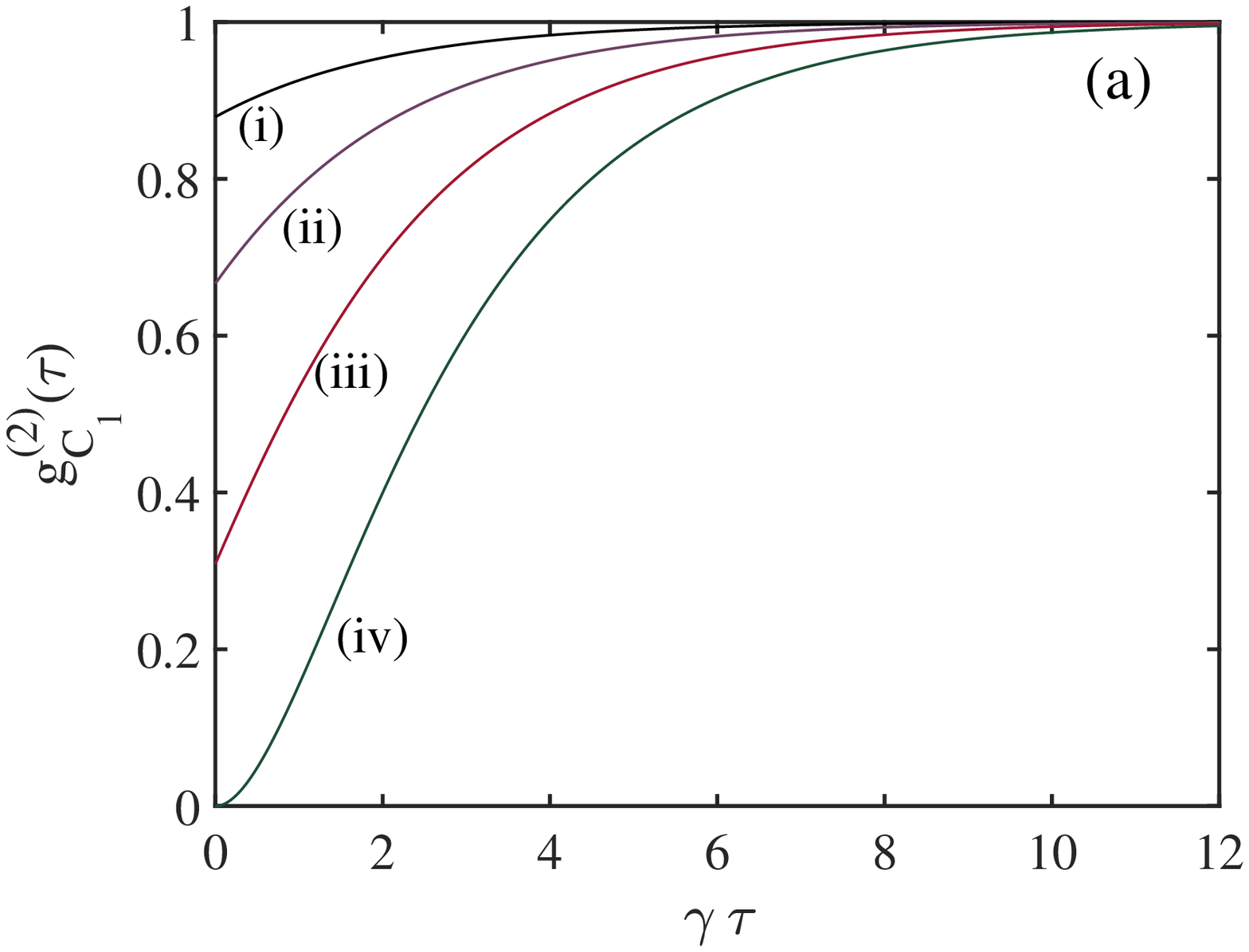}
\includegraphics[width=0.5\textwidth]{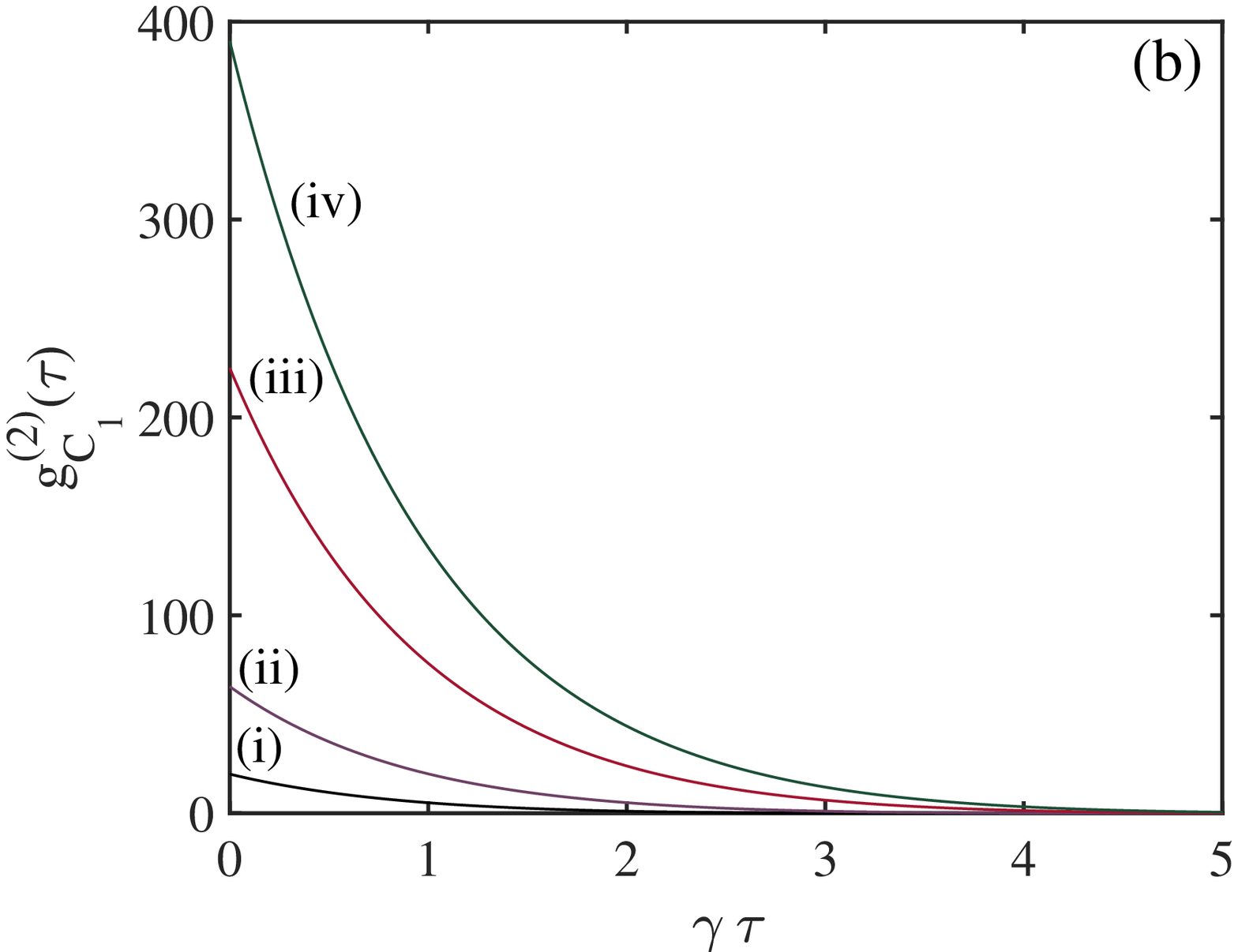}
\end{center}
\caption{{\it Second-order coherence of forwards scattering in the weak-excitation limit.} Second-order correlation function of forwards photon scattering $g^{(2)}_{C_1}(\tau)$ vs the dimensionless delay $\gamma \tau$, calculated using Eq. \eqref{eq:g2apprfinalform}. In {\bf (a)} we depict photon antibunching for $\Gamma=0.2, 0.3, 0.4, 0.5$ as depicted by the curves (i)-(iv), respectively, and in {\bf (b)} the approach to extreme photon bunching for $\Gamma=0.7, 0.75, 0.8, 0.82$ as depicted by the curves (i)-(iv), respectively.}
\label{fig:figurethree}
\end{figure}
\begin{figure}
\begin{center}
\includegraphics[width=0.5\textwidth]{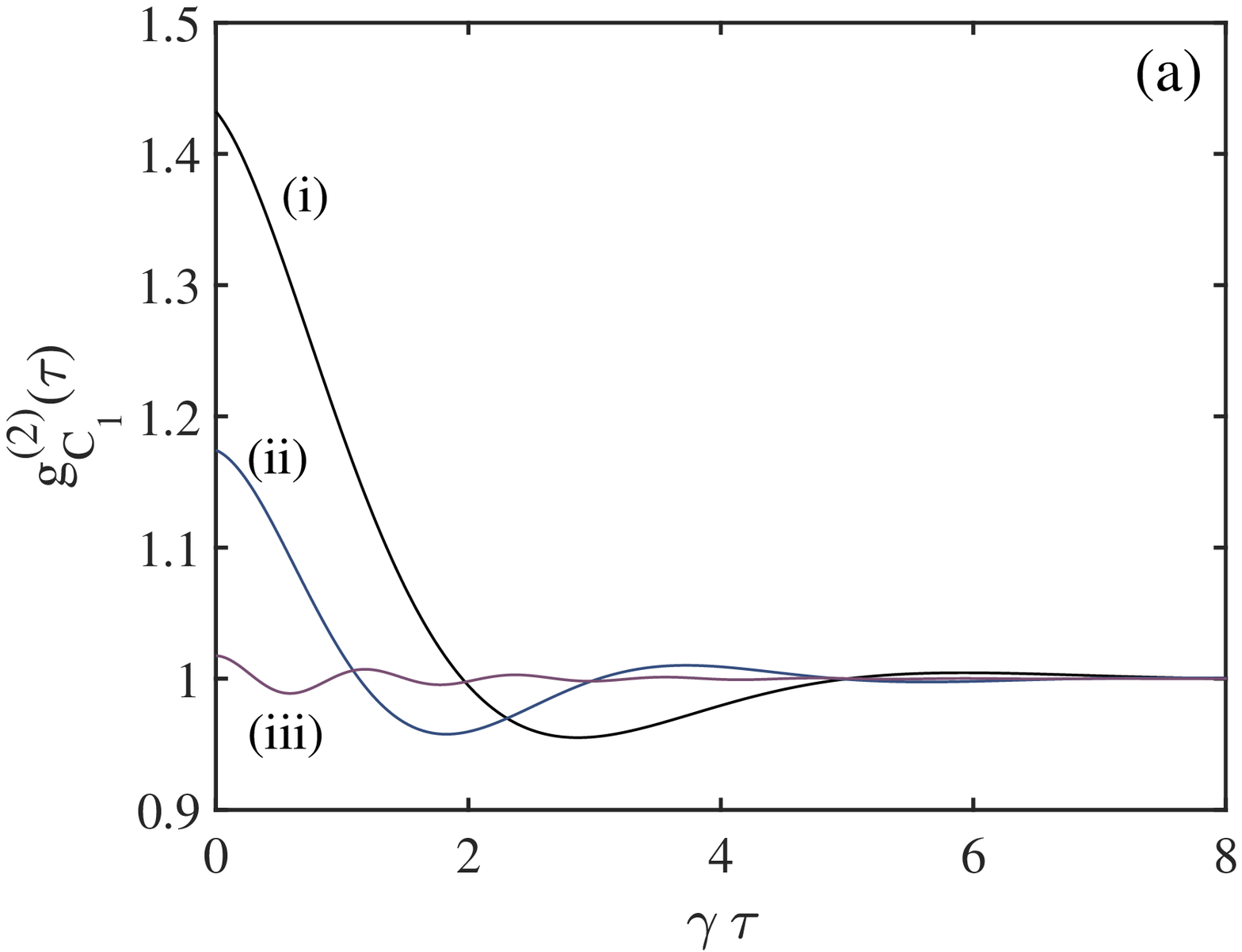}
\includegraphics[width=0.5\textwidth]{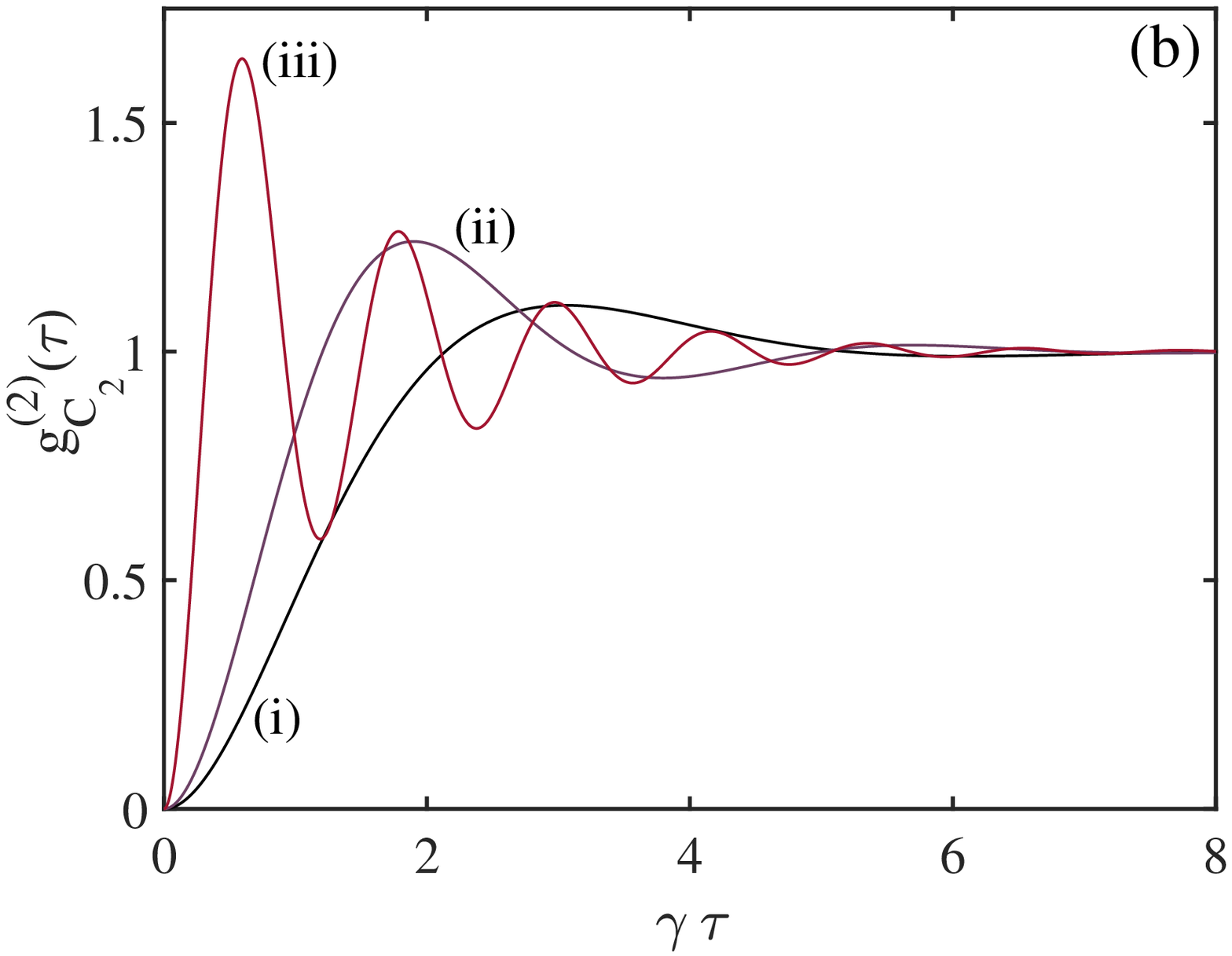}
\end{center}
\caption{{\it Second-order quantum correlations for a weakening coupling to the external atom.} Second-order correlation function of forwards photon scattering $g^{(2)}_{C_1}(\tau)$ in {\bf (a)} and sideways scattering $g^{(2)}_{C_2}(\tau)$ in {\bf (b)} vs the dimensionless delay $\gamma \tau$, for $\gamma/\kappa=0.025, 0.01, 0.001$, as depicted by the curves (i)-(iii), respectively. The correlation functions are derived from the numerical solution of the ME \eqref{eq:ME} through exact diagonalization. The remaining parameters are: $\varepsilon_d/\kappa=0.1$ and $\Gamma=0.7$.}
\label{fig:figurefour}
\end{figure}
This approximation, preserving the purity of the state, is justified only for a weak excitation which guarantees a negligible photon emission probability during the relaxation of the atom back to the steady state. The drive term is accounted for by a non-Hermitian Hamiltonian (retaining only the term proportional to $\sigma_{+}$) preserving the norm of the steady-state reduced density matrix as unity plus a first-order correction in the drive strength. Under this assumption, the second-order correlation function in the forwards direction can be recast in the form
\begin{equation}\label{eq:g2approx}
\begin{aligned}
&g^{(2)}_{C_1}(\tau) \approx (\varepsilon_d/\kappa)^2 (1-\Gamma)^2 \times\\
& \langle\tilde{A}(\tau)|[\varepsilon_d/\kappa+\sqrt{\Gamma \gamma/(4\kappa)}\,\tilde{\sigma}_{+}][\varepsilon_d/\kappa+\sqrt{\Gamma \gamma/(4\kappa)}\,\tilde{\sigma}_{-}]|\tilde{A}(\tau)\rangle,
\end{aligned}
\end{equation}
with initial condition $\ket{\tilde{A}(0)}=\ket{\tilde{A}^{\prime}_{\rm ss}}$. Having eliminated the quantum jumps due to spontaneous emission from the Liouvillian of Eq. \eqref{eq:Liouvillian}, the conditional wavefunction evolves under a non-Hermitian Hamiltonian, obeying the Schr\"{o}dinger equation 
\begin{equation}\label{eq:Schreq}
\frac{d}{d\tau}\ket{\tilde{A}(\tau)}=\left(-\varepsilon_d\sqrt{\frac{\Gamma \gamma}{\kappa}}\sigma_{+}-\frac{\gamma}{2}\sigma_{+}\sigma_{-}\right)\ket{\tilde{A}(\tau)}.
\end{equation} 
We expand the conditional state as $\ket{\tilde{A}(\tau)}=x(\tau)\ket{1} + y(\tau) \ket{2}$, where the complex expansion coefficients $x(\tau),\,y(\tau)$, satisfy the set of coupled linear differential equations
\begin{equation}\label{eq:setofeq}
\frac{dx}{d\tau}=0, \quad \quad \frac{dy}{d\tau}=-\frac{\gamma}{2}\,y-\varepsilon_d \sqrt{\frac{\Gamma\gamma}{\kappa}}\,x.
\end{equation}
The initial conditions should match the conditional state of Eq. \eqref{eq:redstateapp} following the measurement of a photon in the forwards direction, yielding $x(0)=1$ and $y(0)=-2(\varepsilon_d/\kappa)\sqrt{\kappa\Gamma/[\gamma(1-\Gamma)]^2}$. 
\begin{figure}
\begin{center}
\includegraphics[width=0.5\textwidth]{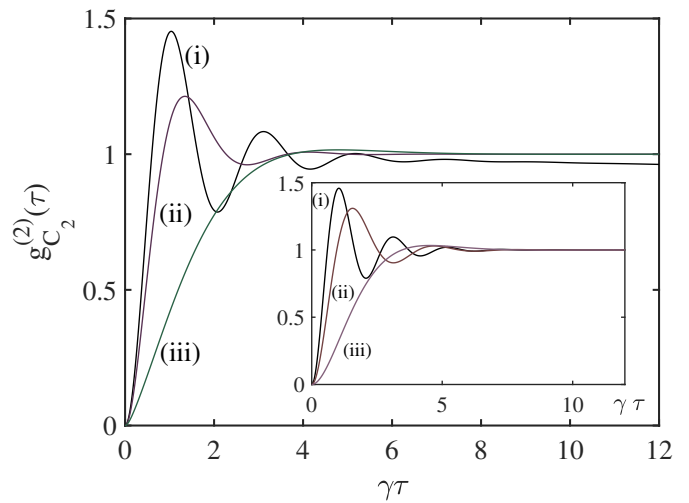}
\end{center}
\caption{{\it Probing the intracavity light-matter coupling strength in the bad-cavity limit.} Second-order intensity correlation function of the sideways scattered light field, $g^{(2)}_{C_2}(\tau)$ vs $\gamma \tau$, extracted from the solution of ME \eqref{eq:ME} with $\gamma_{\rm s}=0$ for increasing $g/\varepsilon_d$ assuming values $0.05$, $1$, and $2.5$ corresponding to the curves (i)-(iii) respectively. The inset depicts the analytical expression of Eq. \eqref{eq:g2side} with the scaled drive amplitude of Eq. \eqref{eq:Yeffresfl}. The remaining parameters are: $\varepsilon_d/\kappa=0.04$, $\Gamma=0.9$, $\gamma/\varepsilon_d=0.0156$ and $\gamma_{\rm s}=0$.} 
\label{fig:figurefive}
\end{figure}

The solution to the set of coupled equations \eqref{eq:setofeq} and their associated initial conditions produces the state
\begin{equation}\label{eq:stateAtau}
\ket{\tilde{A}(\tau)}=\ket{1} - 2\frac{\varepsilon_d}{\kappa} \sqrt{\frac{\kappa\Gamma}{\gamma}}\left(1 + \frac{\Gamma}{1-\Gamma} e^{-(\gamma/2)\tau}\right)\, \ket{2}.
\end{equation} 
Finally, consistent with the initial approximation in Eq. \eqref{eq:state}, which amounts to keeping terms linear in the driving-field amplitude, we write
\begin{equation}\label{eq:aux1}
\begin{aligned}
&\left(\varepsilon_d/\kappa+\sqrt{\Gamma \gamma/(4\kappa)}\,\tilde{\sigma}_{-}\right) \ket{\tilde{A}(\tau)}\\
&\approx (\varepsilon_d/\kappa)\ket{1}-\Gamma(\varepsilon_d/\kappa)[1+\Gamma/(1-\Gamma)\,e^{-(\gamma/2)\tau}]\ket{1}\\
&=\{\varepsilon_d/[\kappa(1-\Gamma)]\}\,\{1-[\Gamma^2/(1-\Gamma)^2]\,e^{-(\gamma/2)\tau}\} \ket{1},
\end{aligned}
\end{equation}
whence, substituting in Eq. \eqref{eq:g2approx}, we finally obtain the {\it second-order correlation function for the forwards emission in the weak-excitation approximation}
\begin{equation}\label{eq:g2apprfinalform}
g^{(2)}_{C_1}(\tau) \approx \left[1-\left(\frac{\Gamma}{1-\Gamma}\right)^2e^{-(\gamma/2)\tau} \right]^2.
\end{equation}
This expression, which is explicitly independent of the driving-field amplitude, agrees with Eq. (41) of \cite{RiceCarmichaelIEEE} with $C\equiv\overline{g}^2/(\kappa\gamma^{\prime})= \Gamma/[2(1-\Gamma)]$, in the correspondence we have introduced in Sec. \ref{sec:cavitycoupledatom}. The transition from photon antibunching to bunching for varying degrees of focusing $\Gamma$ is depicted in Fig. \ref{fig:figurethree}. The function $g^{(2)}_{C_1}(\tau)$ in the weak-coupling limit has a minimum at the delay $\gamma\tau_m=4\ln[\Gamma/(1-\Gamma)]$ (see also Eq. 42 of \cite{RiceCarmichaelIEEE}), which is relevant for $\Gamma \geq 0.5$. In particular, for $\Gamma=0.5$, $\tau_m=0$ and $g^{(2)}_{C_1}(0)=0$, as we can see for curve (iv) in Fig. \ref{fig:figurethree}(a). The occurrence of the minimum value of $g^{(2)}_{C_1}(\tau_m)=0$ for $\tau_m > 0$ evidences the nonclassical character of intensity correlations (see Sec. IV A of \cite{RiceCarmichaelIEEE}). The extreme bunching we observe in Fig. \ref{fig:figurethree}(b), in a clear departure from the second-order coherence of resonance fluorescence, reflects the fact that a measurement in the forwards-scattering channel projects the external atom in its excited state when $\Gamma \to 1$. The excited atom then lets a photon pass through while it is dealing with the one it is about to emit \cite{TrajectoryCascCarmichael, DecoherenceTwostateatom}; as a result, closely-spaced photon pairs are detected, in stark contrast with the scattered field of ordinary resonance fluorescence when the amplitude of the coherent drive is very small. We also note that, owing to the fact that terms proportional to $\sigma_{-} \cdot\sigma_{+}$ \textemdash{destroying} the state purity \textemdash{are} omitted from Eq. \eqref{eq:Liouvillian} in this approximation, $g_{C_1}^{(2)}(\tau)$ is initially proportional to the waiting-time distribution of photon emissions in the forwards direction, $w_{C_1}(\tau)$, a probability distribution over waiting times to the next jump associated with a photon emission, integrated to unity. This proportionality relation holds true for delay times $\gamma\tau$ much smaller than the mean time between jumps (scaled by the atomic lifetime) which, in the limit of a vanishingly weak excitation, extends to infinity (see \cite{Waitingtimedist} and Sec. 13.2.4 of \cite{QO2}). The waiting-time distribution $w_{C_1}(\tau)$, however, ultimately decays to zero at long times (as a probability distribution function), while $g_{C_1}^{(2)}(\tau)$ is asymptotic to unity.

\subsubsection{More on the mapping to a two-level atom inside a coherently driven bad cavity}
\label{subsubsec:mappingbdl}

Following on with our mapping for larger driving strengths and comparing with Eqs. (17) and (23) of \cite{RiceCarmichaelIEEE}, we identify the second-order correlation function of Eq. \eqref{eq:g2fdefinition} with the more involved expression 
\begin{equation}\label{eq:g2badcavity}
\begin{aligned}
&g_{C_1}^{(2)}(\tau)-1=-\frac{8C^2}{[1+Y^2(1+2C)^2]^2}\\
&\times e^{-(3\gamma/4)\tau} \Big\{[1-2C^2 - Y^2(1+2C)^2]\cosh(\delta\tau)\\
&+\frac{\gamma}{4\delta} [1+2C^2-Y^2(1+2C)(5+2C)]\sinh(\delta\tau)\Big\},
\end{aligned}
\end{equation} 
where the parameters $Y$, $\delta$ and $C$ have been defined in Sec. \ref{sec:cavitycoupledatom}, all involving the effective coupling strength $\overline{g}=\sqrt{\kappa\Gamma\gamma/2}$ in the mapping to the bad-cavity configuration. We note, however, that, owing to the unidirectional coupling between the driven cavity \textemdash{whose} resonant mode field remains in a coherent state\textemdash{and} the atom, the validity of Eq. \eqref{eq:g2badcavity} extends {\it beyond} the bad-cavity limit {\it when mapped to the parameters of the cascaded-system setup}. There is no entanglement between the atom and the cavity field due to their {\it direct} coupling \cite{DecoherenceTwostateatom}. This means that a condition of the kind $\kappa/\gamma \gg 1$ \textemdash{the} translation of the bad-cavity rate hierarchy \textemdash{is} {\it not} imposed. Here, the presence of $C$ addresses a different physical mechanism: instead of being the Purcell factor enhancing the spontaneous emission rate to arbitrary values set by the intracavity coupling strength, it promotes the sideways-emission rate to its full $4\pi$ value, $\overline{\gamma}_{\rm s} \to \gamma$.  
\begin{figure}
\begin{center}
\includegraphics[width=0.5\textwidth]{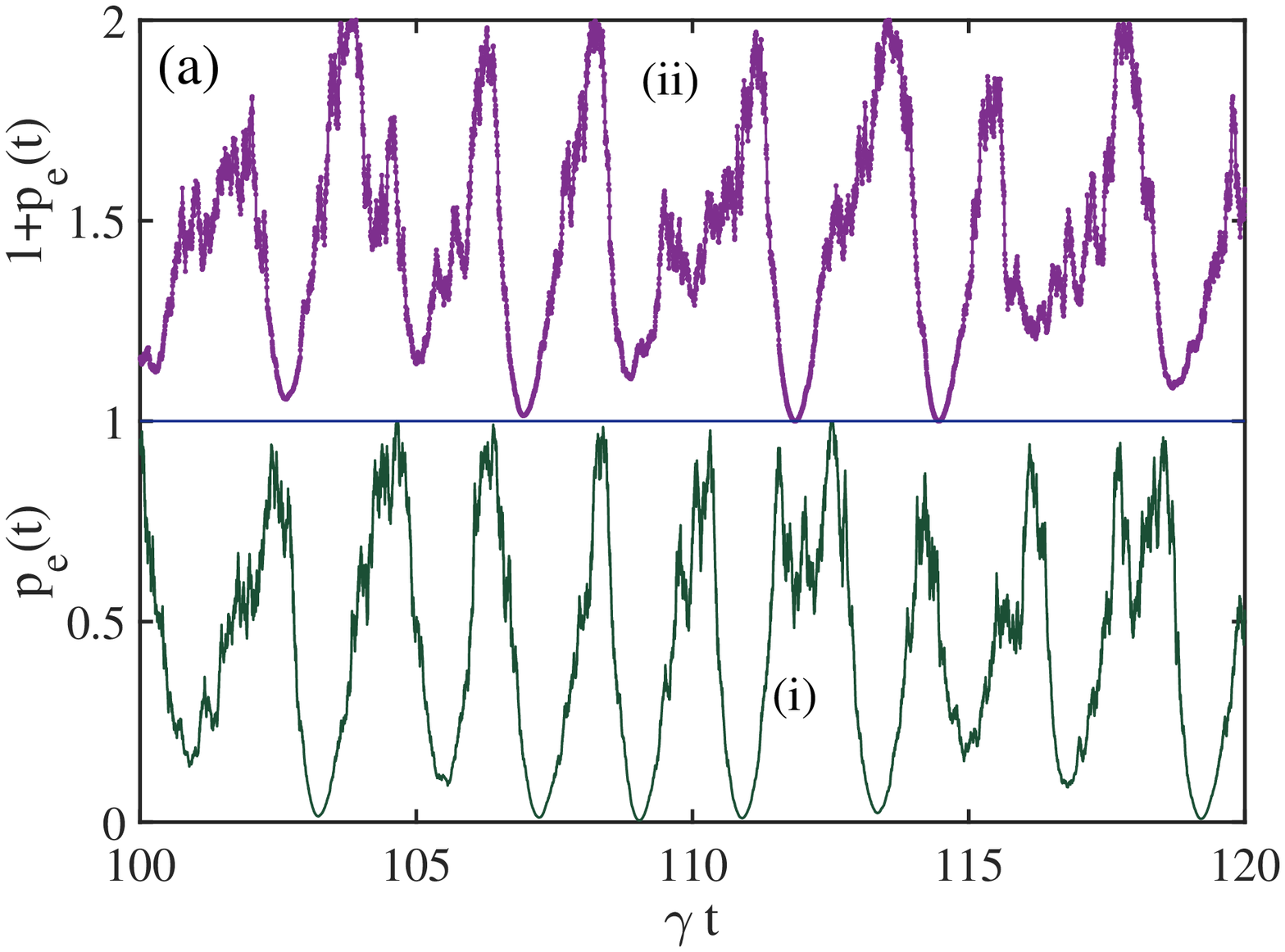}
\includegraphics[width=0.5\textwidth]{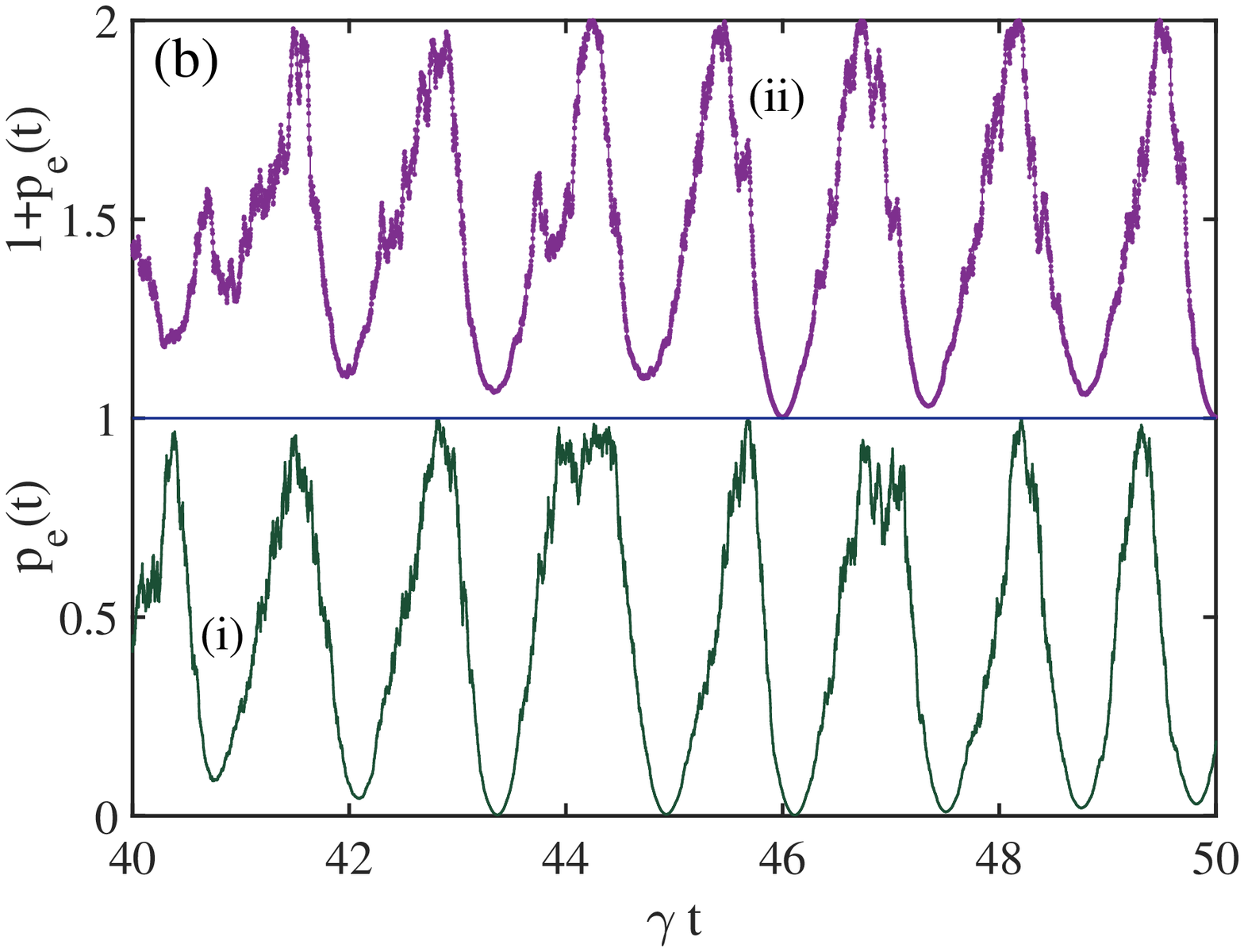}
\end{center}
\caption{{\it Rabi oscillations in single quantum trajectories.} {\bf (a)} Time-dependent probability $p_{\rm e}(t) \equiv 0.5(1+\braket{\sigma_z(t)})$ of finding the atom outside the cavity in the excited state, extracted from a single realization. Curve (i) depicts $p_{\rm e}(t)$ from the unravelling of the full ME \eqref{eq:ME} in the presence of sideways spontaneous emission by the internal two-level atom, with $g/\varepsilon_d=0.5$, $\gamma_{\rm s}/\varepsilon_d=0.25$, $\gamma/\varepsilon_d=0.0156$ and $\Gamma=0.9$. Curve (ii), depicting $1+p_{\rm e}(t)$ (for visual clarity), originates from the reduced ME of ordinary resonance fluorescence. {\bf (b)} Same as in frame (a), but for $\gamma/\varepsilon_d=0.0069$. In all quantum trajectories unravelling the (corresponding) MEs, the same seed to the random-number generator and initial conditions were used; both atoms were initialized in their ground states, and the cavity field in the Fock state $\ket{n=1}$ for the generation of curves (i) in both frames.}
\label{fig:figuresix}
\end{figure}
As for the sideways emission \textemdash{for} the field operator $C_2$ \textemdash{one} employs directly the familiar result from free-space resonance fluorescence, namely (see Appendix and \cite{QO1}),
\begin{equation}\label{eq:g2side}
\begin{aligned}
g_{C_2}^{(2)}(\tau)&=\frac{\braket{\tilde{C}_2^{\dagger}(0)\tilde{C}_2^{\dagger}(\tau)\tilde{C}_2(\tau)\tilde{C}_2(0)}_{\rm ss}}{\braket{\tilde{C_2}^{\dagger}\tilde{C}_2}_{\rm ss}^2}\\
&=1-e^{-(3\gamma/4)\tau}\left(\cosh \delta\tau + \frac{3\gamma}{4\delta} \sinh \delta\tau \right),
\end{aligned}
\end{equation}
with $g_{C_2}^{(2)}(0)=0$ for all values of $\Gamma$. Further evidence on the distinct character of the forwards-emission field alongside its difference from the ordinary resonance fluorescence of sideways emission, is given by the numerically-computed correlation functions of Fig. \ref{fig:figurefour} beyond the weak-excitation limit. The results depicted here coincide with the analytical predictions of Eqs. \eqref{eq:g2badcavity} and \eqref{eq:g2side}. In Fig. \ref{fig:figurefour}(a), the decay of $g^{(2)}_{C_1}(\tau)$ to a minimum below unity [curve (i)] (approaching arbitrarily low values for $\varepsilon_d/\kappa \to 0$ \textemdash{a} nonclassical effect discussed in \cite{RiceCarmichaelIEEE}) is replaced by rapid oscillations of a weak amplitude about unity with increasing $Y$ (as we keep $\varepsilon_d/\kappa$ constant and increase $\kappa/\gamma$) [curve (iii)], in the approach to a coherent-state output. At the same time, $g^{(2)}_{C_2}(\tau)$ shows the expected onset of the free-space resonance fluorescence ringing associated with the pronounced Rabi doublet in the optical and squeezing spectra we have met in Fig. \ref{fig:figuretwo}. For the computation of higher-order correlation functions for the forwards-scattering channel, one employs directly Eqs. 22 and 23 of \cite{RiceCarmichaelIEEE}.

\section{Adiabatic elimination of the intracavity field: free-space resonance fluorescence revisited}
\label{sec:adiabaticelimfield}

We now bring the atom inside the cavity into play in a perturbative fashion, assuming a coupling to the cavity mode with a small but finite strength $g \ll \kappa$. This interaction is considered together with emission from the sides of the cavity at a rate $\gamma_{\rm s}$, adding another dissipation channel to the ME \eqref{eq:ME} via the term $\mathcal{L}[C_3]\tilde{\rho}$, with $C_3=\sqrt{\gamma_{\rm s}}\,\sigma_{1-}$. We focus on a hierarchy of timescales set by the condition $\kappa \gg (\varepsilon_d, g, \gamma_{\rm s}/2)$, such that the cavity relaxes fast to a coherent state with very small amplitude, while the atom inside the cavity remains close to its ground state with a steady-state excitation probability on the order of the scaled drive amplitude squared \textemdash{consistent} with our approximation in Eq. \eqref{eq:state} referring now to the JC oscillator in the bad-cavity limit.

From the ME \eqref{eq:ME} (including $\mathcal{L}[C_3]\tilde{\rho}$), we obtain the following equations of motion for the time-varying averages of the coupled cavity mode field, atomic polarization and inversion of the internal two-level atom (where the tilde on top of the operators signifies the equivalence to a frame rotating with the atomic frequency $\omega_A$):
\begin{subequations}\label{eq:HEofM1}
\begin{align}
&\frac{d\braket{\tilde{a}}}{dt}=-\kappa\braket{\tilde{a}} +g \braket{\tilde{\sigma}_{1-}} +\varepsilon_d,   \label{eq:HEofMa}\\
&\frac{d\braket{\tilde{\sigma}_{1-}}}{dt}=-\frac{\gamma_{\rm s}}{2} \, \braket{\tilde{\sigma}_{1-}} + g \braket{\sigma_{1z}\tilde{a}}, \label{eq:HEofMb}\\
&\frac{d\braket{\tilde{\sigma}_{1+}}}{dt}=-\frac{\gamma_{\rm s}}{2} \, \braket{\tilde{\sigma}_{1+}} + g \braket{{\tilde{a}^{\dagger}\sigma}_{1z}}, \label{eq:HEofMb2}\\
&\frac{d\braket{\sigma_{1z}}}{dt}=-\gamma_{\rm s}(\braket{\sigma_{1z}}+1)-2g(\braket{\tilde{\sigma}_{1+}\tilde{a}}+\braket{\tilde{a}^{\dagger}\tilde{\sigma}_{1-}}), \label{eq:HEofMc}
\end{align}
\end{subequations}
and for the external atom,
\begin{subequations}\label{eq:HEofM2}
\begin{align}
&\frac{d\braket{\tilde{\sigma}_{2-}}}{dt}=-\frac{\gamma}{2}\, \braket{\tilde{\sigma}_{2-}} + \sqrt{\kappa\gamma\Gamma}\,\braket{\sigma_{2z}\tilde{a}}, \label{eq:HEofMd} \\
&\frac{d\braket{\tilde{\sigma}_{2+}}}{dt}=-\frac{\gamma}{2}\, \braket{\tilde{\sigma}_{2+}} + \sqrt{\kappa\gamma\Gamma}\,\braket{\tilde{a}^{\dagger}\sigma_{2z}}, \label{eq:HEofMe} \\
&\frac{d \braket{\sigma_{2z}}}{dt}=-\gamma(\braket{\sigma_{2z}}+1) - 2 \sqrt{\kappa\gamma\Gamma}\,(\braket{\tilde{\sigma}_{2+}\tilde{a}}+\braket{\tilde{a}^{\dagger}\tilde{\sigma}_{2-}}). \label{eq:HEofMf}
\end{align}
\end{subequations}
From the Heisenberg-Langevin equation (12) of \cite{Gardiner1993}, by virtue of the unidirectional coupling between the two cascaded systems and in a frame rotating with $\omega_A$, we obtain \cite{RiceCarmichaelIEEE}
\begin{equation}\label{eq:aadiabaticel}
\tilde{a}(t)=\frac{\varepsilon_d}{\kappa} + \frac{g}{\kappa}\tilde{\sigma}_{1-}(t) + \frac{1}{\kappa}\hat{\xi}(t),
\end{equation}
where $\hat{\xi}(t)$ is the quantum-noise term arising from the interaction of the cavity mode with the field modes of a reservoir (and contains the sum of the corresponding annihilation operators over those modes). Here, we take $\braket{\hat{\xi}(t)}=0$, assuming that the reservoir is in the vacuum state. Substituting the expression of Eq. \eqref{eq:aadiabaticel} for the adiabatically eliminated intracavity field into the equations of motion \eqref{eq:HEofM1} and \eqref{eq:HEofM2}, in which the system operators have been pre-ordered in such a fashion as to make clear that every term involving the reservoir field coming from $\tilde{a}(t)$ is zero, yields the Bloch equations for the two-level atom inside the cavity,
\begin{subequations}\label{eq:HEofMelim1}
\begin{align}
&\frac{d\braket{\tilde{\sigma}_{1-}}}{dt}=-\frac{\gamma_{\rm s}}{2} (1+2C) \, \braket{\tilde{\sigma}_{1-}} + \frac{\overline{Y}}{\sqrt{2}}\,\braket{\sigma_{1z}}, \label{eq:HEofMelima}\\
&\frac{d\braket{\tilde{\sigma}_{1+}}}{dt}=-\frac{\gamma_{\rm s}}{2} (1+2C) \, \braket{\tilde{\sigma}_{1-}} + \frac{\overline{Y}}{\sqrt{2}}\,\braket{\sigma_{1z}}, \label{eq:HEofMelimb}\\
&\frac{d\braket{\sigma_{1z}}}{dt}=-\gamma_{\rm s}(1+2C)(\braket{\sigma_{1z}}+1)\notag\\
&-\frac{\overline{Y}}{2\sqrt{2}}\,(\braket{\tilde{\sigma}_{1+}}+\braket{\tilde{\sigma}_{1-}}), \label{eq:HEofMelimc}
\end{align}
\end{subequations}
and for the external atom,
\begin{subequations}\label{eq:HEofMelim2}
\begin{align}
&\frac{d\braket{\tilde{\sigma}_{2-}}}{dt}=-\frac{\gamma}{2}\, \braket{\tilde{\sigma}_{2-}} + \varepsilon_d\sqrt{\frac{\gamma\Gamma}{\kappa}}\,\braket{\sigma_{2z}}\notag\\
& + g\sqrt{\frac{\gamma\Gamma}{\kappa}}\,\braket{\tilde{\sigma}_{1-}\sigma_{2z}}, \label{eq:HEofMelimd} \\
&\frac{d\braket{\tilde{\sigma}_{2+}}}{dt}=-\frac{\gamma}{2}\, \braket{\tilde{\sigma}_{2+}} + \varepsilon_d\sqrt{\frac{\gamma\Gamma}{\kappa}}\,\braket{\sigma_{2z}}\notag\\
& + g\sqrt{\frac{\gamma\Gamma}{\kappa}}\,\braket{\tilde{\sigma}_{1+}\sigma_{2z}}, \label{eq:HEofMelime} \\
&\frac{d \braket{\sigma_{2z}}}{dt}=-\gamma(\braket{\sigma_{2z}}+1) - 2 \varepsilon_d\sqrt{\frac{\gamma\Gamma}{\kappa}}\,(\braket{\tilde{\sigma}_{2+}}+\braket{\tilde{\sigma}_{2-}})\notag\\
&-2g \sqrt{\frac{\gamma\Gamma}{\kappa}}\,(\braket{\tilde{\sigma}_{1+}\tilde{\sigma}_{2-}} + \braket{\tilde{\sigma}_{1-}\tilde{\sigma}_{2+}} ), \label{eq:HEofMelimf}
\end{align}
\end{subequations}
where in this case there is explicit spontaneous-emission enhancement for the internal two-level system by $(1+2C)$, with the Purcell factor $C=g^2/(\kappa\gamma_{\rm s})$ depending on the strength of the coherent intracavity light-matter interaction.
\begin{figure*}
\begin{center}
\includegraphics[width=\textwidth]{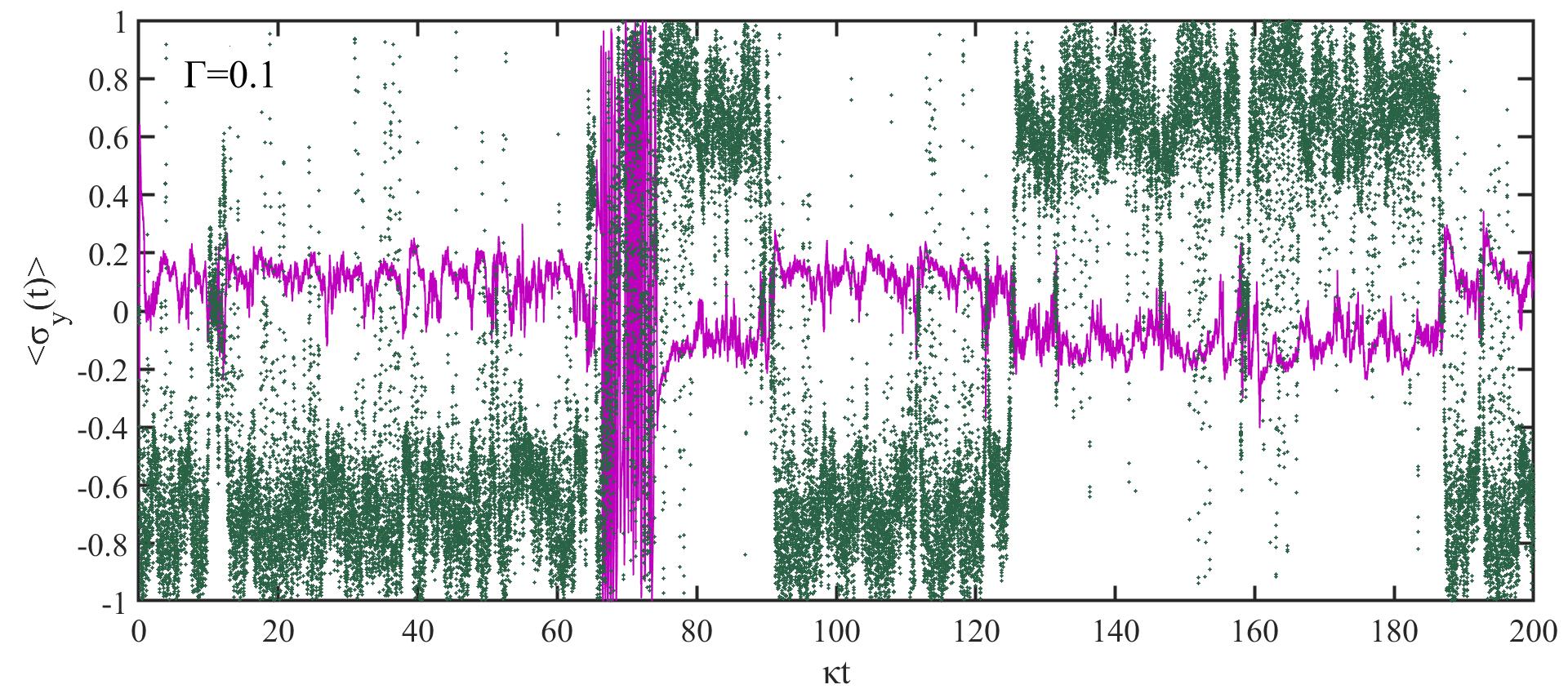}
\includegraphics[width=\textwidth]{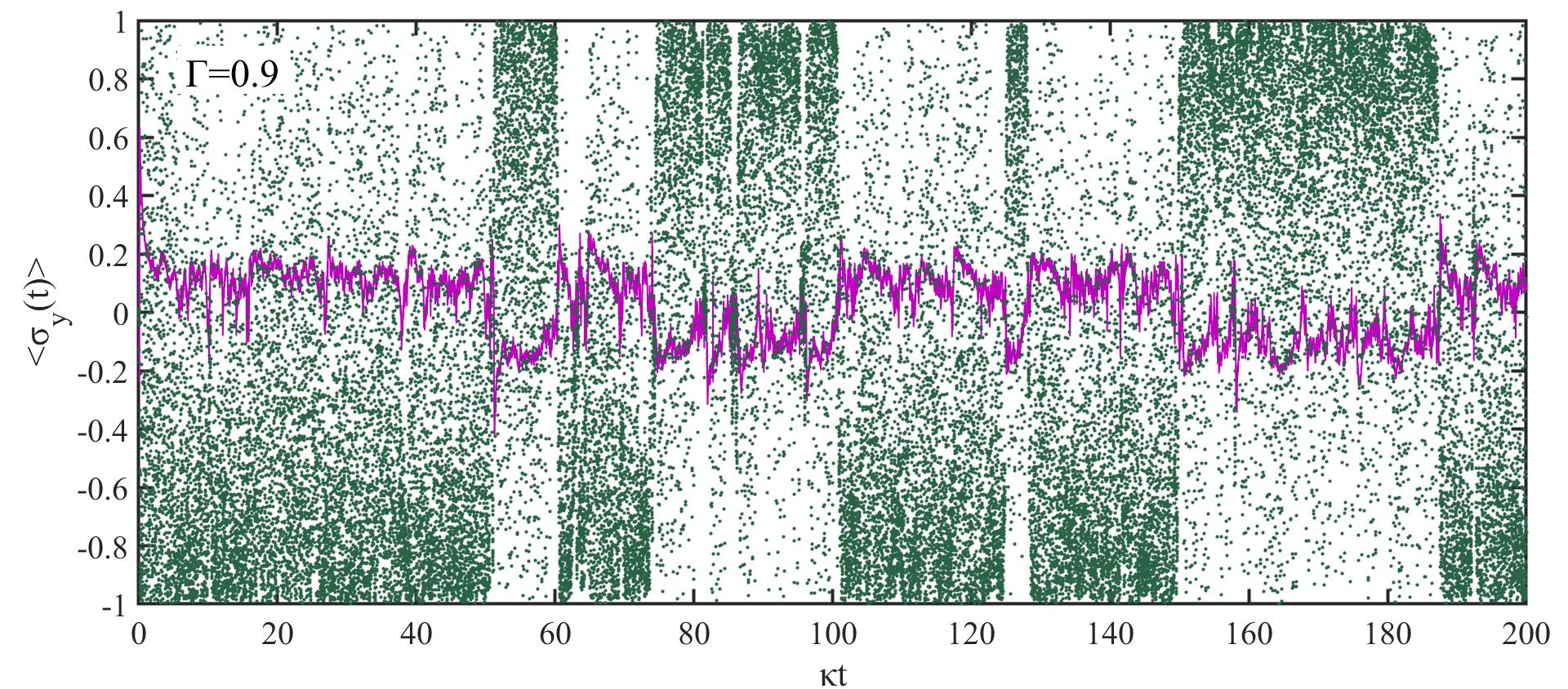}
\end{center}
\caption{{\it Bimodal field driving a two-level atom}. Time-dependent averages $\braket{\sigma_y(t)}$ of the slowly varying $y$-polarization component for the atom inside (in solid purple line) and outside (in green dots) the cavity, extracted from a single quantum trajectory. For the trajectory in the {\bf top} panel, $\Gamma=0.1$, and for the trajectory in the {\bf bottom} panel, $\Gamma=0.9$. The remaining parameters are: $g/\kappa=100$, $\gamma/\kappa=40$ and $\varepsilon_d/g=0.501$. For both quantum trajectories the same seed to the random-number generator and initial conditions were used; both atoms were initialized in their ground states, and the cavity field in the Fock state $\ket{n=1}$.}
\label{fig:figureseven}
\end{figure*}
Equations \eqref{eq:HEofMelima}-\eqref{eq:HEofMelimc} can be solved independently of the quantities pertaining to the external atom, and the steady-state solution follows from the standard treatment of free-space resonance fluorescence (for an atom placed outside the cavity) as \cite{RiceCarmichaelIEEE} 
\begin{equation}\label{eq:firstsetY}
\begin{aligned}
&\braket{\tilde{\sigma}_{1\pm}}_{\rm ss}=-\frac{1}{\sqrt{2}}\frac{\overline{Y}(1+2C)}{(1+2C)^2 + \overline{Y}^2},\\
& \braket{\sigma_{1z}}_{\rm ss}=-\frac{(1+2C)^2}{(1+2C)^2 + \overline{Y}^2},
\end{aligned}
\end{equation}
where $\overline{Y}\equiv 2\sqrt{2}g\varepsilon_d/(\kappa \gamma_{\rm s})$ is the scaled dimensionless drive amplitude. When $\gamma_{\rm s} \to 0$, we can instead write
\begin{equation}\label{eq:firstsetYsmallgamma}
\braket{\tilde{\sigma}_{1\pm}}_{\rm ss}=-\frac{1}{\sqrt{2}}\frac{\overline{Y}^{\prime}}{1 + \overline{Y}^{\prime 2}}, \quad \quad \braket{\sigma_{1z}}_{\rm ss}=-\frac{1}{1 + \overline{Y}^{ \prime 2}},
\end{equation}
with $\overline{Y}^{\prime}=2\sqrt{2} \{g \varepsilon_d/[\kappa \gamma(1+2C)]\} \to \sqrt{2}\varepsilon_d/g$. 

We will now decouple the moments of Eqs. \eqref{eq:HEofMelimd}-\eqref{eq:HEofMelimf} in the mean-field approximation, and seek the steady-state solutions of the modified equations of motion for the external two-level system,
\begin{equation}\refstepcounter{equation}\label{eq:HEofMelim2TWAa}
-\frac{\gamma}{2}\, \braket{\tilde{\sigma}_{2-}}_{\rm ss} + \varepsilon_d\sqrt{\frac{\gamma\Gamma}{\kappa}}\left[1 + \frac{g}{\varepsilon_d}\,\braket{\tilde{\sigma}_{1-}}_{\rm ss}\right]\braket{\sigma_{2z}}_{\rm ss}=0, \tag{\theequation a}
\end{equation}
\begin{equation*}\label{eq:HEofMelim2TWAb} 
-\frac{\gamma}{2}\, \braket{\tilde{\sigma}_{2+}}_{\rm ss} + \varepsilon_d\sqrt{\frac{\gamma\Gamma}{\kappa}}\left[1 + \frac{g}{\varepsilon_d}\,\braket{\tilde{\sigma}_{1+}}_{\rm ss}\right]\braket{\sigma_{2z}}_{\rm ss}=0, \tag{\theequation b}
\end{equation*}
\begin{equation*}\label{eq:HEofMelim2TWAc}
\begin{aligned}
&-\gamma(\braket{\sigma_{2z}}_{\rm ss}+1) - 2\varepsilon_d\sqrt{\frac{\gamma\Gamma}{\kappa}}\left[1 + \frac{g}{\varepsilon_d}\,\braket{\tilde{\sigma}_{1+}}_{\rm ss}\right]\\
&\times \,(\braket{\tilde{\sigma}_{2+}}_{\rm ss}+\braket{\tilde{\sigma}_{2-}}_{\rm ss})=0.
\end{aligned} \tag{\theequation c}
\end{equation*}
The solution to these equations \textemdash{equivalent} to free-space resonance fluorescence \textemdash{is}, as usual, 
\begin{equation}\label{eq:2TWAMF}
\braket{\tilde{\sigma}_{2\pm}}_{\rm ss}=-\frac{1}{\sqrt{2}}\frac{\overline{Y}^{\prime\prime}}{1 + \overline{Y}^{\prime\prime 2}}, \quad \quad \braket{\sigma_{2z}}_{\rm ss}=-\frac{1}{1 + \overline{Y}^{\prime\prime 2}},
\end{equation}
but now with $\overline{Y}^{\prime\prime} \equiv 2\varepsilon_d\sqrt{2\Gamma/(\kappa \gamma)}[1 + (g/\varepsilon_d)\,\braket{\tilde{\sigma}_{1+}}_{\rm ss}]$. For $g/\varepsilon_d \ll 1$ and $\varepsilon_d/\gamma_{\rm s} \ll 1$, one recovers the semiclassical dynamics predicted by Eqs. \eqref{eq:neocl2}. When $\gamma_s \to 0$, we obtain
\begin{equation}\label{eq:Yeffresfl}
\overline{Y}_{\gamma_s \to 0}^{\prime\prime} \equiv 2\varepsilon_d\sqrt{\frac{2\Gamma}{\kappa \gamma}}\left[1-\frac{1}{1+2(\varepsilon_d/g)^2}\right],
\end{equation}
tending to zero for small ratios $\varepsilon_d/g$. The effect of reducing the dimensionless drive amplitude $\overline{Y}_{\gamma_s \to 0}^{\prime\prime}$ when increasing the ratio $g/\varepsilon_d$ is reflected in the intensity correlation functions for sideways scattering of Fig. \ref{fig:figurefive}, where numerical results from the solution of the ME \eqref{eq:ME} are compared to the analytical expression of Eq. \eqref{eq:g2side}, with the appropriate scaled amplitude, taken from Eq. \eqref{eq:Yeffresfl}. The long-time approach of the second-order coherence function for sideways scattering to unity, as depicted in curve (i) of the main plot (for $g/\varepsilon_d \ll 1$), is indicative of quantum correlations built up between the internal atom, coupled to the radiating intracavity field, and the external scatterer. Simulations show that the deviation is larger when $g/\gamma \sim 1$ and for a small intracavity photon number $(\varepsilon_d/\kappa)^2$, which is a sign of departure from the validity of the adiabatic elimination of the intracavity field and the mean-field dynamics of free-space resonance fluorescence. Otherwise, the two sets of curves are in good agreement. 

Prompted by this semiclassical argument, we will now compare the probability to find the external two-level atom in the excited state in single quantum trajectories unravelling the full ME \eqref{eq:ME}, to the resonance fluorescence corresponding to the equations of motion \eqref{eq:HEofMelimd}-\eqref{eq:HEofMelimf}. We assume a factorization of moments whereby the atomic polarization and inversion of the internal two-level atom are kept equal to their steady-state values at all times. In other words, we compare to the solution of a reduced ME where a coherent field drives the external atom, with an amplitude set by the mean-field steady-state operator averages of the atom inside the cavity. The results are depicted in Fig. \ref{fig:figuresix}, where we can observe that a decreasing spontaneous emission rate $\gamma$ brings the increasingly coherent Rabi oscillations in phase with the monitored output of resonance fluorescence. The oscillations depicted in the curves (ii), following from unravelling the ME of the free-space resonance fluorescence, correspond to the steady-state solution of the optical Bloch equations \textemdash{as} given in Eq. \eqref{eq:2TWAMF} \textemdash{for} the scaled drive amplitude $Y^{\prime\prime}$. This amplitude is in turn defined from the parameters employed for the solution of the full ME \eqref{eq:ME}, unravelled when producing the curves (i). In the meanwhile, the inversion for the atom inside the cavity (in the cascaded configuration) remains virtually fixed at its steady-state value $\braket{\sigma_{1z}}_{\rm ss} \approx -0.95$, yielding a very small probability of finding the atom in the excited state \textemdash{in} line with the weak-excitation limit of Sec. \ref{subsubsec:weakElimit} \textemdash{as} we expect in the bad-cavity limit of the JC interaction we are here considering. By lowering substantially the photon loss rate $2\kappa$ with respect to the coupling strength $g$ and considering the limit of zero spontaneous emission, we access the strong-coupling regime which is not amenable to perturbation theory: there, modifications occur at the level of coherent quantum dynamics due to the significant JC nonlinearity (Sec. 13.3 of \cite{QO2}) transmitted through to the monitored output. 
\begin{figure*}
\begin{center}
\includegraphics[width=\textwidth]{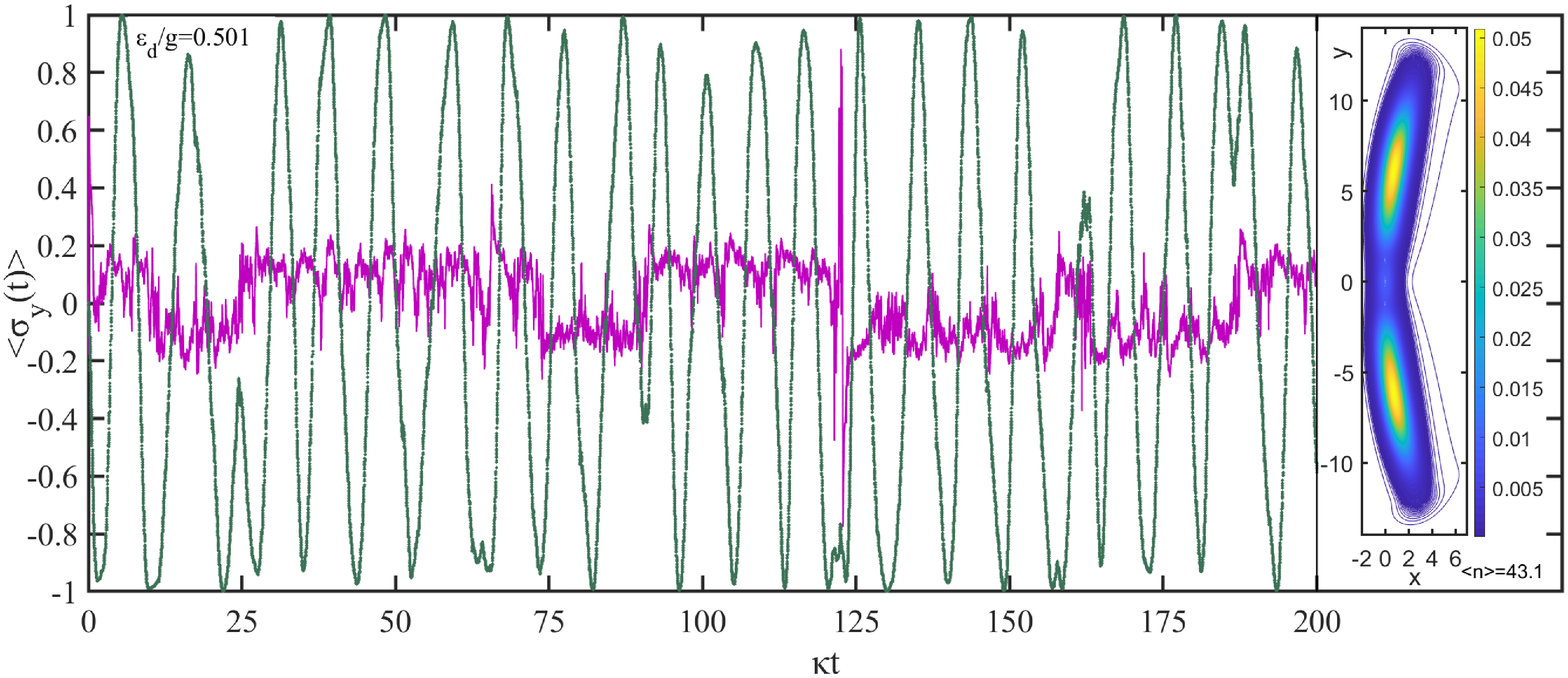}
\includegraphics[width=\textwidth]{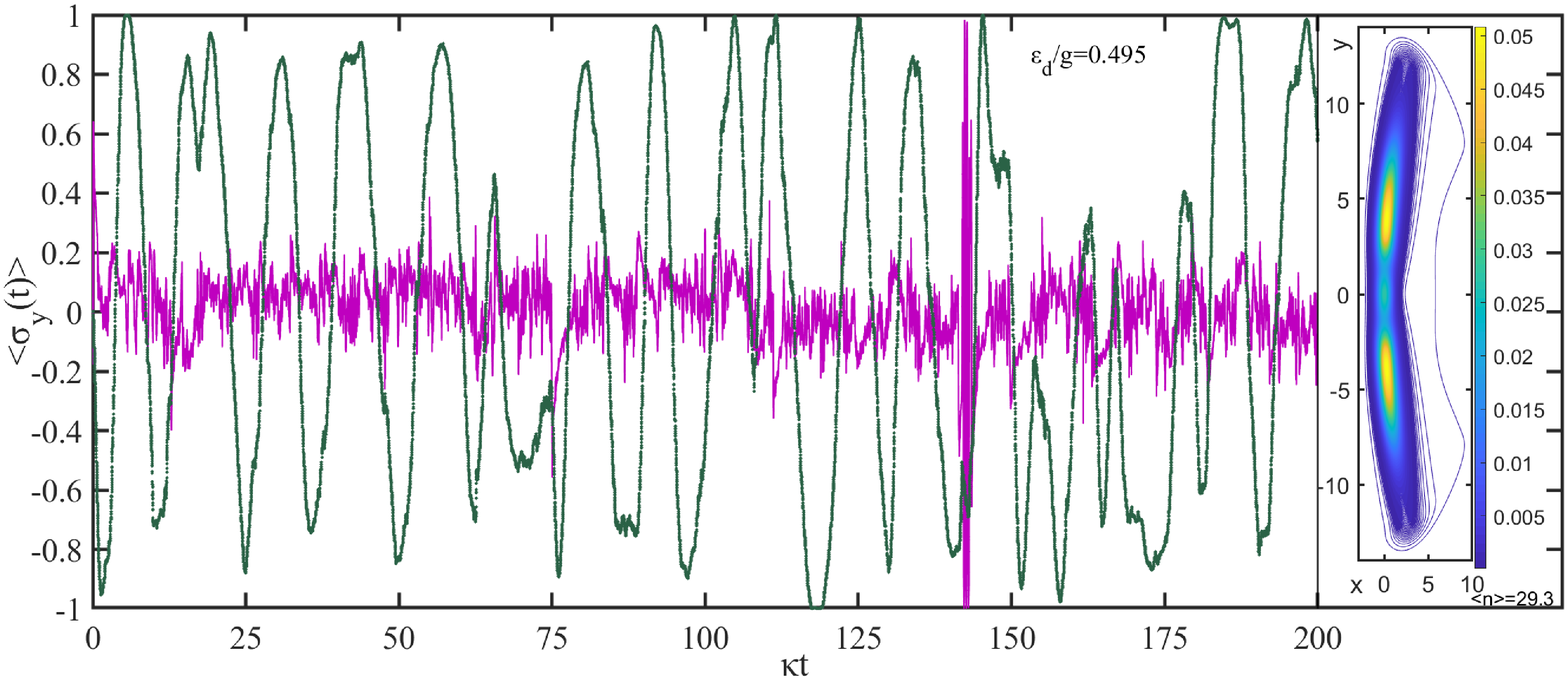}
\end{center}
\caption{{\it Driving the external atom above and below the critical point of symmetry breaking.} Time-dependent averages $\braket{\sigma_y(t)}$ of the slowly varying $y$-polarization component for the atom inside (in solid purple line) and outside (in green dots) the cavity, extracted from a single quantum trajectory. For the quantum trajectory depicted in the {\bf top} panel, $\varepsilon_d/g=0.501$, and for the trajectory in the {\bf bottom} panel, $\varepsilon_d/g=0.495$. The insets depict the quasi-probability distribution $Q(x+iy)$ of the intracavity field, with the corresponding steady-state photon number given underneath. The remaining parameters are: $g/\kappa=100$, $\gamma/\kappa=0.004$ and $\Gamma=0.95$. As in Fig. \ref{fig:figureseven}, for both quantum trajectories the same seed to the random-number generator and initial conditions were used; both atoms were initialized in their ground states, and the cavity field in the Fock state $\ket{n=1}$.}
\label{fig:figureeight}
\end{figure*}

\section{Quantum-fluctuation bimodal switching driving an external two-level atom}
\label{sec:phasebist}

The strong-coupling regime is defined by the condition $g/\kappa \gg 1$; here, $g/2$ and $\varepsilon_d$ are of the same order of magnitude, while we also assume that the internal atom is not radiatively coupled to the modes of the vacuum field ($\gamma_{\rm s}=0$), carrying on from Sec. \ref{sec:modelandsetup}. By doing so we reach the limit of ``zero system size'' $\gamma_{\rm s}^2/(8g^2) \to 0$, bringing spontaneous dressed-state polarization into play (see Sec. 16.3 of \cite{QO2} and \cite{Alsing_1991}). The output channel reflects the collapse of the quasi-energy spectrum at the critical point $\varepsilon_d=g/2$ in the associated second-order dissipative quantum phase transition \cite{CarmichaelPhotonBlockade}. Above threshold, $\varepsilon_d \geq g/2$, the neoclassical (see Sec. IIC of \cite{CarmichaelPhotonBlockade}) steady-state intracavity field is bimodal according to the expression \cite{Alsing_1991, DiracJC}
\begin{equation}\label{eq:bimodalityalpha}
\tilde{\alpha}_{\rm ss}=\frac{\varepsilon_d}{\kappa}\left[1-\left(\frac{g}{2\varepsilon_d}\right)^2\right] \pm i \frac{g}{2\kappa}\sqrt{1-\left(\frac{g}{2\varepsilon_d}\right)^2},
\end{equation}
identifying a complex-conjugate pair of state amplitudes. In the strong-coupling limit, where $g/\kappa \gg 1$, and sufficiently away from the critical point, the mean-field solutions \eqref{eq:steadystateMF} for the two-level atom outside the cavity are (for $\Gamma\kappa/\gamma \sim 1$, guaranteeing $|Y| \gg 1$)
\begin{equation}\label{eq:bistatomout1}
\tilde{\beta}_{2, \,{\rm ss}}\approx-\frac{1}{\sqrt{2}} \frac{Y}{|Y|^2}=-\frac{1}{4}\sqrt{\frac{\gamma}{\Gamma\kappa}}\,\frac{\alpha_{\rm ss}}{|\alpha_{\rm ss}|^2},
\end{equation}
yielding
\begin{equation}
\begin{aligned}
&\tilde{\beta}_{2, \,{\rm ss}}\approx -\frac{1}{4}\sqrt{\frac{\gamma}{\Gamma\kappa}} \left[\sqrt{1-\left(\frac{g}{2\varepsilon_d}\right)^2}\pm i\frac{g}{2\varepsilon_d}\right]\\
&\times\left\{\left(\frac{\varepsilon_d}{\kappa}\right)^2\left[1-\left(\frac{g}{2\varepsilon_d}\right)^2\right]\right\}^{-1/2},
\end{aligned}
\end{equation}
while
\begin{equation}\label{eq:bistatomout2}
\zeta_{2, \,{\rm ss}}\approx -\frac{1}{|Y|^2}=-\frac{\gamma}{8\Gamma\kappa}\left\{\left(\frac{\varepsilon_d}{\kappa}\right)^2\left[1-\left(\frac{g}{2\varepsilon_d}\right)^2\right]\right\}^{-1}.
\end{equation}
For the two-level atom inside the cavity, the corresponding quantities read [see Eqs. \eqref{eq:neocl2b}-\eqref{eq:neocl2c} and \eqref{eq:conservation} of Sec. \ref{sec:modelandsetup}]
\begin{equation}\label{eq:bistatomin}
\tilde{\beta}_{1, \,{\rm ss}}=\pm i \frac{\tilde{\alpha}_{\rm ss}}{2|\tilde{\alpha}_{\rm ss}|}=-\frac{g}{4\varepsilon_d} \pm i \frac{1}{2}\sqrt{1-\left(\frac{g}{2\varepsilon_d}\right)^2}, \quad \zeta_{1, \,{\rm ss}}=0.  
\end{equation}
Defining $\lambda \equiv (g/2\varepsilon_d)^2$, we observe that at the critical point, $\lambda_c=1$, both the field amplitudes $\tilde{\alpha}_{\rm ss}$ and the atomic polarization states $\tilde{\beta}_{(1,2),\,{\rm ss}}$ display a pitchfork-like bifurcation. The complex order parameter $\tilde{\alpha}_{\rm ss}$ of Eq. \eqref{eq:bimodalityalpha} points to a scaling of the form $(\lambda-\lambda_c)^{1/2}$, identifying a critical exponent equal to $1/2$. However, the moduli of the external polarization, $|\tilde{\beta}_{2,\,{\rm ss}}|$, and inversion, $\zeta_{2,\,{\rm ss}}$, scale instead as $(\lambda-\lambda_c)^{-1/2}$ and $(\lambda-\lambda_c)^{-1}$, respectively, for large excitation amplitudes $Y$ away from the critical point, while the modulus of the internal polarization, $|\tilde{\beta}_{1,\,{\rm ss}}|$ is equal to $1/2$ above the critical point, and the internal inversion $\zeta_{1,\,{\rm ss}}$ remains fixed at zero.

Quantum-fluctuation bistable switching above threshold, stabilizing the mean-field states as attractors \cite{Alsing_1991}, is depicted in Fig. \ref{fig:figureseven} for low and high degree of focusing to the external atom, and a steady-state average intracavity photon number $\braket{n}_{\rm ss}\equiv \braket{a^{\dagger}a}_{\rm ss} \approx 43$. The imaginary part of the polarization-operator average has opposite signs for the two atoms, $1$ and $2$, as correctly predicted by the semiclassical Eqs. \eqref{eq:bistatomout1} and \eqref{eq:bistatomin}. The two metastable states with conjugate polarization have a lifetime that significantly exceeds the $\kappa^{-1}$-timescale set by dissipation. Since $\gamma/\kappa \gg 1$, one cannot distinguish individual Rabi oscillations (for the external atom) during the lifetime of each of the two states of polarization with opposite imaginary parts. Increasing the degree of focusing from $\Gamma=0.1$ to $\Gamma=0.9$ influences the particular realization of quantum-fluctuation switching in the bistable JC oscillator, with switching events occurring at different positions, as we can observe when comparing the two panels of Fig. \ref{fig:figureseven}. We observe that the external atom responds to the bistable switching even for a low degree of focusing. When comparing the two individual realizations, we also note the disappearance of the time interval characterized by intense decoherence (for $60 \leq \kappa t \leq 80$) in Fig. \ref{fig:figureseven}(a) when focusing to the external atom is stronger [Fig. \ref{fig:figureseven}(b)]; in the latter case, the Bloch vector explores larger regions of the unit sphere. 

In Fig. \ref{fig:figureeight}, we observe the distinction between driving the external two-level atom by a state that fluctuates and by a state with nonzero mean-field amplitude $\tilde{\alpha}_{\rm ss}$ slightly below and above the critical point, respectively. Quantum-fluctuation switching between the two semiclassical solutions of the external atom, as seen in both frames, is simultaneously accompanied by phase jumps of the cavity-field amplitude. Hence, the two subsystems become phase correlated via radiative coupling to the same reservoir. Here, $\gamma/\kappa \ll 1$ (resulting in a much larger value of $Y$ than the one used in Fig. \ref{fig:figureseven}); therefore, individual Rabi oscillations are visible, with a period which is comparable to the lifetime of the metastable states. Switching events to a different metastable state \textemdash{and} accordingly to a different excitation ladder of the JC spectrum \citep{Alsing_1991, CarmichaelPhotonBlockade} \textemdash{in} the top panel of Fig. \ref{fig:figureeight}, are correlated with a clear disruption of the Rabi-oscillation cycles. This is a disruption of the phase arising from the switching of the drive-field phase, in contrast to a disruption due to spontaneous emission \textemdash{which} resets the atomic oscillation to the ground state \textemdash{seen} in trajectories of regular resonance fluorescence. For the bottom panel of Fig. \ref{fig:figureeight}, the scaled amplitude of the field driving external atom is $Y=0$, since the cavity is driven by a field with an amplitude below its threshold value (whence $\tilde{\alpha}_{\rm ss}=0$). Nevertheless, coherent oscillations in the imaginary part of the atomic-polarization average, though visibly more distorted, can still be discerned, together with their disruptions, correlated with transitions to a different metastable state. We need to emphasize here that this is a regime of pronounced fluctuations, being around the critical point of a second-order quantum phase transition in zero dimensions, where a departure from the mean-field predictions is to be expected when one monitors directly the output of the bistable oscillator (see Sec. 16.3.6 of \cite{QO2}). In fact, the $Q$-function for the cavity-field distribution in Fig. \ref{fig:figureeight} evidences two maxima for complex-conjugate amplitudes, before the appearance of the expected mean-field bifurcation; this suggests a conditional evolution roughly of the type described in Eqs. (50-52) of \cite{Alsing_1991}, affected and monitored by the external two-level atom, in spite of driving below threshold.

\section{Concluding discussion and future work}

In this work, we have derived analytical results for the statistics of the forwards and sideways emission channel by means of a mapping to an atom {\it inside} the coherently driven cavity coupled to the supported resonant mode with a strength determined by the dissipation rates of the initial cascaded-system configuration. For this purpose, we at first set to zero the coupling strength between the cavity and the two-level atom comprising the JC oscillator, developing a treatment which relied on several well-known results from ordinary resonance fluorescence and the bad-cavity limit of QED. We then brought progressively the JC dynamics into play, reflected in the nonlinearity of the nonclassical light emanating from the first subsystem coupled to an atomic scatterer. Through a succession of coupled equations of motion for the two cascaded subsystems, we employed a semiclassical and numerical investigation to compare the solution of the full ME with free-space resonance fluorescence for the atom-scatterer lying outside the cavity. The latter is driven by an effective field whose amplitude is determined by the steady-state polarization and inversion averages of the atom inside the cavity.

By promoting the intracavity coupling strength between the two constituents of the JC oscillator we have eventually moved to a regime where the quantum nature of the field driving the external atom cannot be reduced to a mean-field or perturbative description. The output of the bistable oscillator in the region of the critical point, forming part of the forwards-scattering channel, involves actively both coupled quantum degrees of freedom in the JC interaction (which is {\it not} the case when $g/\kappa \ll 1, \gamma_{\rm s}/\kappa \ll 1$); it is in a state of pronounced quantum fluctuations subject to a conditional evolution which actively involves the external two-level emitter. Away from the critical point, where the quasi-energy spectrum collapses, such an output can be approximated by a mixed state with an equal representation of the two quasi-coherent states of phase bimodality \cite{Alsing_1991}. In the absence of monitoring by the external atom ($\gamma=0$), we have also found that the conditional evolution of phase bistability in the coupled degrees of freedom is different from the trajectories depicted in Fig. \ref{fig:figureseven} for the same seed to the random-number generator. Therefore, our analysis has been largely based upon the coherence properties of resonance fluorescence and the associated Rabi oscillations, with reference to the different statistics of forwards and sideways scattering for an external atom coupled to the coherently driven cavity; such a disparity arises due to the interference with the coherent cavity output field, and varies with the degree of focusing. Disruptions in these coherent oscillations, emerging on approaching the critical point, are correlated with the switching events realizing phase bistability in the individual trajectories unravelling the evolution of the full system density matrix (source plus target).

Let us pause here to comment a little bit further on the mapping to the bad-cavity limit, which, as we have already pointed out, effectively amounts to placing the external atom {\it inside} the cavity and dealing with JC dynamics, even if perturbatively. In the cascaded-system configuration, the forwards dipole scattering is made significant by strongly focusing the cavity output onto the external emitter so that a significant fraction of the $4\pi$ solid angle seen by the atom is occupied by the incoming mode. In this case, the drive field must be mode-matched to the dipole mode that naturally arises in the coupling of a dipole transition to the electromagnetic field in free space \textemdash{this} remains a considerable challenge on the experimental front. There is no Purcell enhancement involved in generating the mode overlap and, in that sense, our mapping is rather formal. Nevertheless, the effective ``enhancement'' of the sideways spontaneous emission rate to its full $4\pi$ value, points us to the ME \eqref{eq:MEatom} in which a classical field is driving the target atom (a central message of \cite{DecoherenceTwostateatom}) where the trace over the cavity field has worked out to yield exactly the total emission rate $\gamma$.

Following the resurgence of interest in the emission properties of a target driven by an explicitly quantum source, an immediate extension of our work could touch upon driving the external atom with the output of a JC oscillator in the regime of photon blockade, probing its persistence in the ``thermodynamic limit'' of strong coupling, where the scale parameter grows with the coupling strength. Here, the output is a stream of distinct photon assemblies, corresponding to the multiphoton resonances responsible for the blockade, and comprises a source of manifestly nonclassical light (depending on the strength of the driving, both bunching and antibunching may occur \textemdash{see} Sec. 3.3 of \cite{Shamailov2010}). The experiment would then focus on the radiation properties of such a source in an environment containing the external atom-target \textemdash{another} quantum nonlinear oscillator \textemdash{monitoring} the composite conditional evolution via the two channels of the distributed forwards and sideways emission.\\

\begin{acknowledgments}
Th. K. M. is grateful to H. J. Carmichael for instructive discussions and guidance, as well as to R. Guti\'{e}rrez-J\'{a}uregui for helpful comments and suggestions on the manuscript. Both authors acknowledge financial support by the Swedish Research Council (VR) in conjunction with the Knut and Alice Wallenberg foundation (KAW).
\end{acknowledgments}

\bibliography{bibliography}

\onecolumngrid

\appendix*
\section{Auxiliary expressions from resonance fluorescence}

In this Appendix, we calculate first and second-order correlation functions for the source field of the atom outside the cavity, which we then use to derive expressions for the incoherent spectra and spectra of squeezing of the two channels in Secs. \ref{subsec:incoherentspectrum} and \ref{subsec:spectrumsq}, respectively, as well as their second-order coherence properties in Sec. \ref{subsubsec:mappingbdl}.

The expectation value of the fluctuations in $\boldsymbol{s}^{\top} \equiv (\tilde{\sigma}_{-}, \tilde{\sigma}_{+}, \sigma_z)$ \textemdash{defined} in a frame rotating with $\omega_A$\textemdash{populating} the vector of fluctuation operators  
\begin{equation}
\Delta \boldsymbol{s} \equiv \begin{pmatrix}
\Delta \tilde{\sigma}_{-} \\ \Delta \tilde{\sigma}_{+} \\ \Delta\sigma_z
\end{pmatrix} \equiv \begin{pmatrix}
\tilde{\sigma}_{-} \\ \tilde{\sigma}_{+} \\ \sigma_z
\end{pmatrix}-\begin{pmatrix}
\braket{\tilde{\sigma}_{-}}_{\rm ss} \\ \braket{\tilde{\sigma}_{+}}_{\rm ss} \\ \braket{\sigma_z}_{\rm ss}
\end{pmatrix},
\end{equation}
evolves in time according to
\begin{equation}
\frac{d}{dt}\braket{\Delta \boldsymbol{s}}=\boldsymbol{M}\braket{\Delta \boldsymbol{s}}.
\end{equation}
The matrix $\boldsymbol{M}$ is provided by the Bloch equations, extracted from the ME of resonance fluorescence,
\begin{equation}
\frac{d{\rho}}{dt}=-i\frac{1}{2}\omega_A[\sigma_z, \rho]-\frac{\gamma Y}{2\sqrt{2}}[\sigma_{+}e^{-i\omega_A t} - \sigma_{-}e^{i\omega_A t}, \rho] + \frac{\gamma}{2}(2\sigma_{-}\rho \sigma_{+}-\sigma_{+}\sigma_{-}\rho-\rho \sigma_{+}\sigma_{-}),
\end{equation}
as
\begin{equation}
\boldsymbol{M} \equiv \begin{pmatrix}
-\gamma/2 & 0 & \sqrt{\kappa\gamma\Gamma}\alpha_{\rm ss} \\
0 & -\gamma/2 & \sqrt{\kappa\gamma\Gamma}\alpha_{\rm ss} \\
-2\sqrt{\kappa\gamma\Gamma}\alpha_{\rm ss} & -2\sqrt{\kappa\gamma\Gamma}\alpha_{\rm ss} & -\gamma
\end{pmatrix}=-\frac{\gamma}{2}\begin{pmatrix}
1 & 0 & -Y/\sqrt{2} \\
0 & 1 &- Y/\sqrt{2} \\
\sqrt{2}Y & \sqrt{2}Y & 2
\end{pmatrix}.
\end{equation}
The quantum regression formula, then, dictates the evolution of the first-order correlation function, $\braket{\Delta\tilde{\sigma}_{+}(0)\Delta\boldsymbol{s}(\tau)}_{\rm ss}$,
\begin{equation}\label{eq:quantumregr}
\frac{d}{d\tau}\braket{\Delta\tilde{\sigma}_{+}(0)\Delta\boldsymbol{s}(\tau)}_{\rm ss}=\boldsymbol{M}\braket{\Delta\tilde{\sigma}_{+}(0)\Delta\boldsymbol{s}(\tau)}_{\rm ss}. 
\end{equation}
The initial conditions are given by the steady-state values
\begin{equation}\label{eq:initial}
\braket{\Delta\tilde{\sigma}_{+}\Delta\boldsymbol{s}}_{\rm ss}=\begin{pmatrix}
\frac{1}{2}(1+\braket{\sigma_z}_{\rm ss})-\braket{\tilde{\sigma}_{+}}_{\rm ss}\braket{\tilde{\sigma}_{-}}_{\rm ss}\\
-\braket{\tilde{\sigma}_{+}}_{\rm ss}^2\\
-\braket{\tilde{\sigma}_{+}}_{\rm ss}(1+\braket{\sigma_z}_{\rm ss})
\end{pmatrix}=\frac{1}{2}\frac{Y^2}{(1+Y^2)^2}\begin{pmatrix}
Y^2 \\ -1 \\ \sqrt{2} Y
\end{pmatrix}
\end{equation}
The formal solution to Eq. \eqref{eq:quantumregr} is given by
\begin{equation}\label{eq:formalsol}
\braket{\Delta\tilde{\sigma}_{+}(0)\Delta\boldsymbol{s}(\tau)}_{\rm ss}=\boldsymbol{S}^{-1}\exp(\boldsymbol{\lambda}\tau)\boldsymbol{S}\braket{\Delta\tilde{\sigma}_{+}\Delta\boldsymbol{s}}_{\rm ss},
\end{equation}
where $\boldsymbol{\lambda} \equiv \boldsymbol{S}\boldsymbol{M}\boldsymbol{S}^{-1}={\rm diag}(\lambda_1, \lambda_2, \lambda_3)$, with $\lambda_1=-\gamma/2$ and $\lambda_{2,3}=-3\gamma/4\pm\delta$, is a diagonal matrix formed by the eigenvalues of $\boldsymbol{M}$. Here, the shift $\delta$ captures the dependence of the eigenvalues on the driving strength, and is defined as $\delta \equiv (\gamma/4)\sqrt{1-8Y^2}$. There is one special point where $\boldsymbol{M}$ becomes non-diagonalizable, namely when $\delta=0$ [or $Y=1/(2\sqrt{2})$]. This is a so-called {\it exceptional point}, at which two of the eigenvalues, $\lambda_2$ and $\lambda_3$, coalesce. At that point, these two eigenvalues switch from purely real (relaxing response) to complex (decaying and oscillatory response), which coincides with the formation of the Mollow triplet depicted in Fig. \ref{fig:figuretwo}. Since $\boldsymbol{M}$ is a non-Hermitian matrix, its left and right eigenvectors are in principle not equivalent; the rows of $\boldsymbol{S}$ are then populated by the left eigenvectors of $\boldsymbol{M}$, while the columns of $\boldsymbol{S}^{-1}$ are populated by the right eigenvectors of $\boldsymbol{M}$. The right eigenvector corresponding to the eigenvalue $\lambda_1$ is $\boldsymbol{e}_1=(1/\sqrt{2})(1, -1, 0)^{\mathsf{T}}$ (which is also equal to the transpose of the corresponding left eigenvector). The remaining right and left eigenvectors corresponding to the eigenvalues $\lambda_{2,3}$ assume the form $\boldsymbol{e}_2=c_2(1,1,A_2)^{\mathsf{T}}$, $\boldsymbol{e}_3=c_3(1,1,A_3)^{\mathsf{T}}$ and $\boldsymbol{e}_2^{\prime}=c_2^{\prime}(1,1,A_2^{\prime})$, $\boldsymbol{e}_3^{\prime}=c_3^{\prime}(1,1,A_3^{\prime})$, respectively. The coefficients featuring in the third components of the eigenvectors read:
\begin{equation}\label{eq:coeff}
A_2=\left(\delta-\frac{\gamma}{4}\right)\frac{2\sqrt{2}}{Y\gamma}, \quad A_3=-\left(\delta+\frac{\gamma}{4}\right)\frac{2\sqrt{2}}{Y\gamma}, \quad A_2^{\prime}=\left(\frac{\gamma}{4}-\delta \right)\frac{\sqrt{2}}{Y\gamma}, \quad A_3^{\prime}=\left(\frac{\gamma}{4}+\delta \right)\frac{\sqrt{2}}{Y\gamma}.
\end{equation}
Following then the prescription, we write Eq. \eqref{eq:formalsol} in the form
\begin{equation*}
\braket{\Delta\tilde{\sigma}_{+}(0)\Delta\boldsymbol{s}(\tau)}_{\rm ss}=\begin{pmatrix}
1/\sqrt{2} & c_2 & c_3 \\
-1/\sqrt{2} & c_2 & c_3 \\
0 & A_2 c_2 & A_3 c_3 
\end{pmatrix} \exp(\boldsymbol{\lambda}\tau) \begin{pmatrix}
1/\sqrt{2} & -1/\sqrt{2} & 0 \\
c_2^{\prime} & c_2^{\prime} & A_2^{\prime}c_2^{\prime} \\
c_3^{\prime} & c_3^{\prime} & A_3^{\prime} c_3^{\prime}
\end{pmatrix}\braket{\Delta\tilde{\sigma}_{+}\Delta\boldsymbol{s}}_{\rm ss},
\end{equation*}
where 
\begin{equation*}
\exp(\boldsymbol{\lambda}\tau)=\begin{pmatrix}
e^{-(\gamma/2)\tau} & 0 & 0 \\
0 & e^{-(3\gamma/4-\delta)\tau} & 0 \\
0 & 0 & e^{-(3\gamma/4+\delta)\tau}
\end{pmatrix}.
\end{equation*}
The orthonormality of right and left eigenvectors produces the system of equations 
\begin{subequations}\label{eq:orthconditions}
\begin{align}
&A_2 A_2^{\prime} c_2 c_2^{\prime} + A_3 A_3^{\prime} c_3 c_3^{\prime}=1, \label{eq:cond1}\\
&A_2^{\prime} c_2 c_2^{\prime} + A_3^{\prime} c_3 c_3^{\prime}=0. \label{eq:cond2}
\end{align}
\end{subequations}
Solving the above system of equations yields
\begin{equation*}
c_2 c_2^{\prime}=[A_2^{\prime}(A_2-A_3)]^{-1}=(1/4)[1+\gamma/(4\delta)], \quad \quad A_2^{\prime} c_2 c_2^{\prime}=-A_3^{\prime} c_3 c_3^{\prime}=Y\gamma/(4\sqrt{2}\delta), \quad c_3 c_3^{\prime}=(1/4)[1-\gamma/(4\delta)].
\end{equation*}
Then, for the various first-order correlation functions we obtain (see also \cite{Mollow1969, KochanCarmichael1994})
\begin{equation}\label{eq:corr1}
\begin{aligned}
\braket{\Delta\tilde{\sigma}_{+}(0)\Delta\tilde{\sigma}_{-}(\tau)}_{\rm ss}&=\frac{1}{2}\frac{Y^2}{(1+Y^2)^2}\Bigg\{\left[\frac{1}{2}Y^2-\frac{1}{2}(-1)\right]e^{-(\gamma/2)\tau}\\
& + \left[c_2 c_2^{\prime} Y^2 + c_2 c_2^{\prime} (-1) + A_2^{\prime}c_2 c_2^{\prime}(\sqrt{2}Y)\right]e^{-(3\gamma/4-\delta)\tau}\\
&+\left[c_3 c_3^{\prime} Y^2 + c_3 c_3^{\prime} (-1) + A_3^{\prime}c_3 c_3^{\prime}(\sqrt{2}Y)\right]e^{-(3\gamma/4+\delta)\tau} \Bigg\}\\
&=\frac{1}{4}\frac{Y^2}{1+Y^2}e^{-(\gamma/2)\tau}-\frac{1}{8}\frac{Y^2}{(1+Y^2)^2}\left[1-Y^2 + \left(\frac{\gamma}{4\delta}\right)(1-5Y^2)\right]e^{-(3\gamma/4-\delta)\tau} \\
&-\frac{1}{8}\frac{Y^2}{(1+Y^2)^2}\left[1-Y^2 - \left(\frac{\gamma}{4\delta}\right)(1-5Y^2)\right]e^{-(3\gamma/4+\delta)\tau},
\end{aligned}
\end{equation} 
\begin{equation}\label{eq:corr2}
\begin{aligned}
\braket{\Delta\tilde{\sigma}_{+}(0)\Delta\tilde{\sigma}_{+}(\tau)}_{\rm ss}&=\frac{1}{2}\frac{Y^2}{(1+Y^2)^2}\Bigg\{\left[-\frac{1}{2}Y^2+\frac{1}{2}(-1)\right]e^{-(\gamma/2)\tau}\\
& + \left[c_2 c_2^{\prime} Y^2 + c_2 c_2^{\prime} (-1) + A_2^{\prime}c_2 c_2^{\prime}(\sqrt{2}Y)\right]e^{-(3\gamma/4-\delta)\tau}\\
&+\left[c_3 c_3^{\prime} Y^2 + c_3 c_3^{\prime} (-1) + A_3^{\prime}c_3 c_3^{\prime}(\sqrt{2}Y)\right]e^{-(3\gamma/4+\delta)\tau} \Bigg\}\\
&=-\frac{1}{4}\frac{Y^2}{1+Y^2}e^{-(\gamma/2)\tau}-\frac{1}{8}\frac{Y^2}{(1+Y^2)^2}\left[1-Y^2 + \left(\frac{\gamma}{4\delta}\right)(1-5Y^2)\right]e^{-(3\gamma/4-\delta)\tau} \\
&-\frac{1}{8}\frac{Y^2}{(1+Y^2)^2}\left[1-Y^2 - \left(\frac{\gamma}{4\delta}\right)(1-5Y^2)\right]e^{-(3\gamma/4+\delta)\tau},
\end{aligned}
\end{equation} 
\begin{equation}\label{eq:corr3}
\begin{aligned}
\braket{\Delta\tilde{\sigma}_{+}(0)\Delta\sigma_{z}(\tau)}_{\rm ss}&=\frac{1}{2}\frac{Y^2}{(1+Y^2)^2}\Bigg\{\left[A_2c_2 c_2^{\prime} Y^2 + A_2c_2 c_2^{\prime} (-1) + A_2A_2^{\prime}c_2 c_2^{\prime}(\sqrt{2}Y)\right]e^{-(3\gamma/4-\delta)\tau}\\
&+\left[A_3c_3 c_3^{\prime} Y^2 + A_3c_3 c_3^{\prime} (-1) + A_3 A_3^{\prime}c_3 c_3^{\prime}(\sqrt{2}Y)\right]e^{-(3\gamma/4+\delta)\tau} \Bigg\}\\
&=\frac{1}{2\sqrt{2}}\frac{Y^3}{(1+Y^2)^2}\Bigg\{\left[1-\left(\frac{\gamma}{4\delta}\right)(2-Y)\right]e^{-(3\gamma/4-\delta)\tau} + \left[1+\left(\frac{\gamma}{4\delta}\right)(2-Y)\right]e^{-(3\gamma/4+\delta)\tau}\Bigg\}.
\end{aligned}
\end{equation} 
We also note that
\begin{equation*}
\braket{\Delta\tilde{\sigma}_{+}(0)\Delta\boldsymbol{s}(\tau)}_{\rm ss}=\begin{pmatrix}
\braket{\tilde{\sigma}_{+}(0)\tilde{\sigma}_{-}(\tau)}_{\rm ss} \\ \braket{\tilde{\sigma}_{+}(0)\tilde{\sigma}_{+}(\tau)}_{\rm ss} \\
\braket{\tilde{\sigma}_{+}(0)\sigma_{z}(\tau)}_{\rm ss}  
\end{pmatrix}-\begin{pmatrix}
\braket{\tilde{\sigma}_{+}}_{\rm ss}\braket{\tilde{\sigma}_{-}}_{\rm ss} \\
\braket{\tilde{\sigma}_{+}}_{\rm ss}^2\\
\braket{\tilde{\sigma}_{+}}_{\rm ss} \braket{\sigma_z}_{\rm ss}
\end{pmatrix}=\begin{pmatrix}
\braket{\tilde{\sigma}_{+}(0)\tilde{\sigma}_{-}(\tau)}_{\rm ss} \\ \braket{\tilde{\sigma}_{+}(0)\tilde{\sigma}_{+}(\tau)}_{\rm ss} \\
\braket{\tilde{\sigma}_{+}(0)\sigma_{z}(\tau)}_{\rm ss}  
\end{pmatrix}-\frac{Y}{(1+Y^2)^2}\begin{pmatrix}
2Y\\
2Y\\
1/\sqrt{2}
\end{pmatrix}.
\end{equation*}
We now calculate the (normalized) second-order correlation function,
\begin{equation}
g_{\rm ss}^{(2)}(\tau)=\left[\braket{\sigma_{+}\sigma_{-}}_{\rm ss}+\lim_{\tau \to \infty}\braket{\sigma_{+}(0)\sigma_z(\tau)\sigma_{-}(0)}_{\rm ss}\right]^{-1}\,\left[\braket{\sigma_{+}\sigma_{-}}_{\rm ss}+\braket{\sigma_{+}(0)\sigma_z(\tau)\sigma_{-}(0)}_{\rm ss}\right]
\end{equation}
requiring the third component of the vector $\braket{\sigma_{+}(0)\boldsymbol{s}(\tau)\sigma_{-}(0)}_{\rm ss}$. Using once more the quantum regression theorem, this vector evaluates to
\begin{equation*}
\braket{\sigma_{+}(0)\boldsymbol{s}(\tau)\sigma_{-}(0)}_{\rm ss}=\braket{\sigma_{+}\sigma_{-}}_{\rm ss}\braket{\boldsymbol{s}(\tau)}_{\rho(0)=|1\rangle \langle 1|}=\frac{1}{2}\frac{Y^2}{1+Y^2}\braket{\boldsymbol{s}(\tau)}_{\rho(0)=|1\rangle \langle 1|},
\end{equation*}
giving
\begin{equation}\label{eq:2ndcorr2z}
\braket{\sigma_{+}(0)\sigma_z(\tau)\sigma_{-}(0)}_{\rm ss}=-\frac{1}{2}\frac{Y^2}{(1+Y^2)^2} \left[1+Y^2e^{-(3\gamma/4)\tau} \left(\cosh \delta \tau + \frac{3\gamma}{4\delta} \sinh \delta \tau \right) \right],
\end{equation}
and
\begin{equation}\label{eq:2ndcorr2pm}
\begin{aligned}
\braket{\sigma_{+}(0)\tilde{\sigma}_{\pm}(\tau)\sigma_{-}(0)}_{\rm ss}&=-\frac{1}{2\sqrt{2}}\frac{Y^3}{(1+Y^2)^2} \left[1-e^{-(3\gamma/4)\tau} \left(\cosh \delta \tau + \frac{3\gamma}{4\delta} \sinh \delta \tau \right) \right]\\
&-\frac{1}{\sqrt{2}}\frac{Y^3}{1+Y^2}e^{-(3\gamma/4)\tau} \frac{\gamma}{4\delta}\sinh \delta \tau.
\end{aligned}
\end{equation}
Finally,
\begin{equation}
g_{\rm ss}^{(2)}(\tau)=\left(2\braket{\sigma_{+}\sigma_{-}}_{\rm ss}\right)^{-1}\,\left[1+\braket{\sigma_z(\tau)}_{\rho(0)=|1\rangle \langle 1|}\right]=1-e^{-(3\gamma/4)\tau}\left(\cosh \delta\tau + \frac{3\gamma}{4\delta} \sinh \delta\tau \right).
\end{equation}

\vspace{5mm}

\begin{center}
*****
\end{center}

\end{document}